\documentclass[11pt,a4paper]{article}
\usepackage{geometry,multirow,fancyhdr,graphicx,pdflscape}
\usepackage{float,setspace,color,parskip,lscape,amsmath}
\usepackage[labelsep=period,justification=justified,singlelinecheck=false]{caption}
\usepackage{authblk,tabulary,eurosym,epstopdf}
\usepackage[position=top]{subfig}
\usepackage[apaciteclassic]{apacite}
\graphicspath{{Figures/}}
\begin{document}

\title{Moving to Communicate, Moving to Interact: \\ Patterns of Body Motion in Musical Duo Performance}
\author[1]{Laura Bishop}
\author[1,2]{Carlos Cancino-Chac{\'o}n}
\author[3]{Werner Goebl}
\affil[1]{Austrian Research Institute for Artificial Intelligence (OFAI), Vienna, Austria}
\affil[2]{Institute of Computational Perception, Johannes Kepler University Linz, Austria}
\affil[3]{Dept. of Music Acoustics, University of Music and Performing Arts Vienna, Austria}
\date{}
\maketitle

\vspace{-8cm}
\small Bishop, L., Cancino-Chac\'on, C., and Goebl, W. (2019). Moving to communicate, moving to interact: Patterns of body motion in musical duo performance. \emph{Music Perception, 37,} 1-25. This version is a final preprint of the paper prepared by the authors. \normalsize
\vspace{6cm}

\section*{Abstract}
Skilled ensemble musicians coordinate with high precision, even when improvising or interpreting loosely-defined notation. Successful coordination is supported primarily through shared attention to the musical output; however, musicians also interact visually, particularly when the musical timing is irregular. This study investigated the performance conditions that encourage visual signalling and interaction between ensemble members. Piano and clarinet duos rehearsed a new piece as their body motion was recorded. Analyses of head movement showed that performers communicated gesturally following held notes. Gesture patterns became more consistent as duos rehearsed, though consistency dropped again during a final performance given under no-visual-contact conditions. Movements were smoother and interperformer coordination was stronger during irregularly-timed passages than elsewhere in the piece, suggesting heightened visual interaction. Performers moved more after rehearsing than before, and more when they could see each other than when visual contact was occluded. Periods of temporal instability and increased familiarity with the music and co-performer seem to encourage visual interaction, while specific communicative gestures are integrated into performance routines through rehearsal. We propose that visual interaction may support successful ensemble performance by affirming coordination throughout periods of temporal instability and serving as a social motivator to promote creative risk-taking.

\bigskip
\textbf{Keywords:} ensemble performance, musical interaction, communication, movement, gesture

\section{Introduction}
Musical notation provides performers with instructions that have to be interpreted as part of the performance process (i.e., translated into expressive musical output). As these score instructions form an often imprecise template for performance, different performers are likely to interpret them in different ways. In the Western art music tradition (as in many traditions), performers are expected to demonstrate originality in their interpretations while simultaneously respecting relatively strict genre conventions. This challenge -- devising an interpretation that is original but still within the bounds of the tradition -- is not trivial, particularly for ensembles, who must converge on a single interpretation even though performers' individual interpretations might differ. While expressive performance in Western art music has been the subject of much study, the question of how ensembles coordinate shared interpretations remains a neglected issue. Attention to this issue has implications for our understanding of performance in different contexts, including scored, scripted, and entirely oral traditions.

How then do ensemble musicians negotiate a shared musical interpretation and coordinate a cohesive performance in real-time? Study of professional ensembles suggests that they use a combination of communication strategies, including verbal discussion (usually only during rehearsals), visual signalling via gestures and facial expressions, and, critically, shared attention to the combined audio output \cite{Bishop2015,Davidson2012,Fulford2018,Seddon2009,Williamon2002}. Such strategies are readily observed in laboratory settings (e.g., gesturing, as in \citeNP{Davidson2012}) and measurable under experimental manipulations (e.g., reliance on audio signals, as in \citeNP{Bishop2015}). However, these communication strategies do not provide a complete picture of how ensemble coordination is achieved.

In the literature, a distinction has been made between communicative behaviours that involve controlled, one-way transfers of ideas, and those that involve two-way, often pre-reflective, mutually-constructive exchanges \cite{King2017}. We will refer to the former as ``signalling'' and the latter as ``interaction''. \citeA{King2017} suggest that signalling happens primarily in rehearsal settings, while performance involves mostly interaction. Both signalling and interaction processes can be considered deliberate to the extent that they are partially accessible to the performer's attention and control during planned (i.e., goal-directed) coordination. However, some aspects of interaction seem to occur unintentionally (e.g., error correction to maintain coordinated timing, \shortciteNP<see>{Goebl2009,Konvalinka2010,vanderSteen2013}).

As music performance is a multimodal task, signalling and interaction processes span auditory, visual, and motor modalities. Audio is musicians' primary means of interaction, and coordination disintegrates when a reliable exchange of audio signals (as sounded musical output) is not possible \cite{Bishop2015}. Several studies have shown that musicians visually attend to each other's body gestures when the tempo is unstable or irregular, \cite{Bishopsubmitted}, at points of abrupt tempo change \cite{Kawase2014}, and at piece onset \cite{Bishop2018}. In such cases, audio cues may be insufficient for performers to maintain temporal coordination. Recent research has also shown that performers choose to watch each other even during periods of regular timing, and do so increasingly as their familiarity with the music and their co-performers improves \cite{Bishopsubmitted}. Thus, visual signalling and interaction may be more useful to performers than research has so far been able to demonstrate. 

The aim of the present study was to look for evidence of signalling and interaction between duo musicians at the level of body movement. By manipulating musical structure and visual contact and comparing performances given before, during, and after a rehearsal session, we investigated how variability in musical timing, access to visual cues, and familiarity with the music and each other influence performers' movement characteristics and between-performer relationships in movement patterns.

\subsection{Communicative gestures in music performance}
In the music performance literature, ``gestures'' refer to body movements that are inherently meaningful  \shortcite{Dahl2010,Demos2014,Jensenius2010}. Musicians' performance gestures are continuous, coarticulated, and generally multifunctional. A sequence of movements may serve sound-producing and communicative functions simultaneously, for example. Much of the research on the communicative functions of performance gestures has focused on the audience perspective. Performers' body movements have been shown to influence audience perceptions of musical timing \cite{Schutz2007} and expression in audiovisual stimuli \shortcite{Behne2011,Platz2012,Vuoskoski2016}. 

Another line of research has focused more specifically on characteristics of communicative gestures \cite{Bishop2018b}. For example, a study by \shortciteA{Teixeira2015} measured recurrence in clarinet bell movement across performances and identified points in the music where within-performer recurrence was high. \citeA{Thompson2012} examined body sway in pianists, and found that differences in expressive intent corresponded to differences in movement magnitude at points of structural significance. Studies investigating the temporal alignment of expressive gestures with sounded expressive nuances have shown that gesture onsets tend to precede the onset of sounded nuances, suggesting that periods of embodied preparation may help performers realize their intended sound output \shortcite{Livingstone2009,Wanderley2005}. 

Some other research has considered how individual differences among audience members influence their perception of performers' gestures. Music perception is an active, embodied process that makes use of audience members' own motor systems and action-related knowledge \shortcite{Jeannerod2003,Lahav2012,Taylor2014}. Motor expertise is known to improve observers' understanding of familiar gestures and their ability to predict the outcomes of those gestures \shortcite{Aglioti2008,CalvoMerino2006,Petrini2009}. 

Together, these lines of study contribute to an understanding of how performers' gestures -- though often idiosyncratic -- relate in predictable ways to piece structure and, drawing on a shared understanding of genre-specific expressive conventions, communicate aspects of the performer's interpretation to the audience.

This literature provides an important backdrop to the current study, as ensemble musicians are audiences to each other's playing as well as contributors to the collective musical output. The gesture features that influence audience members' perceptions of performance expression, if visible to other ensemble members, may influence their evolving perceptions of the collective output as well. In the current study, we expected to observe performers influencing each other via their body gestures, both in situations where explicit visual signalling is known to occur (e.g., during periods of high temporal variability) and in passages where visual signalling is unnecessary for temporal coordination. 

In the following sections, we discuss the role of body gestures in visual signalling and interaction. Importantly, we also explain how evidence of visual signalling and interaction was expected to manifest in the context of the current study.

\subsection{Visual communication in small ensembles: Cueing gestures}
Signalling via body gestures happens most prominently in the case of large ensembles that are led by a conductor. Conductors' gestures communicate expressive nuances as well as timing. Perceptions of expressivity have been shown to relate positively to gesture amplitude, velocity, and variance \cite{Luck2010}. Tempo is communicated via gesture periodicity. Musicians align their performed beats with peaks in gesture acceleration, rather than with specific points in gesture trajectories \cite{Luck2009}, and they synchronize more precisely with smooth gestures than with gestures that are high in jerk \shortcite{Woellner2012}.

Instrumentalists in small ensembles occasionally exchange conductor-like gestures to clarify the timing of upcoming beats and facilitate synchronization. Most notably, this occurs at piece onset, at re-entry points, and following long pauses \cite{Bishop2015,Bishop2018}. In a study designed to map the direction of musicians' visual attention during duo performance, \citeA{Bishopsubmitted} found that partner-directed gaze peaked just prior to piece onset, during the final notes of the piece, and during an unmetered passage with long held notes. Duo performers thus seem prompted to monitor each other visually when uncertain of whether temporal coordination will be successful, perhaps because they expect that their partner will give a gestural cue. Just as for conductor gestures, beats in instrumentalists' cueing gestures correspond to acceleration peaks and are clearer to observers when gestures are smooth and large in magnitude \cite{Bishop2017}. 

These observations of intentional gestural signalling support the hypothesis, discussed in the joint action literature, that interpersonal coordination requires collaborating group members to predict each other's individual intentions accurately \shortcite{Keller2007,Vesper2010}. The cueing gestures that musicians exchange in moments of uncertainty should serve to clarify their intended timing and improve their own predictability. People have been shown to make themselves more predictable by modifying their behaviour across both musical and non-musical joint action tasks \shortcite{Hart2014,Vesper2016}.

In the current study, we looked for evidence of visual signalling at moments of heightened temporal irregularity. Performers were expected to exchange deliberate cueing gestures (e.g., head nods) at these moments. In line with prior research \cite{Williamon2002}, we expected that some of these cueing gestures would be integrated into the performance plan over the course of rehearsals and repeated across performances. 

\subsection{Visual interaction in small ensembles: Gesture coordination}
The process of continuously predicting and aligning individual intentions, ultimately with the aim of achieving a shared intended output, may not fully explain how music ensembles coordinate at such a high level. In particular, this explanation does not seem to account for the degree of flexibility and spontaneity that we see in skilled musicians' performances \cite{Bishop2018a,Moran2014,Schiavio2017}. At its peak, ensemble performance is thought to be ``emergent'', meaning that expressive nuances arise that were not part of the original performance plan (as suggested by the score, implied by genre conventions, and/or established during rehearsals), and cannot be attributed to any one ensemble member \cite{Cochrane2017,Hart2014}. Instead, new ideas are thought to emerge in real-time as ensemble members respond to fluctuations in each other's behaviour and continuously evaluate and re-evaluate their collective, gradually unfolding output.

Interaction between performers -- including at the level of their body movement -- is thought to be central to the real-time coordination of newly emerging ideas. A number of studies have documented coordination in the gestures performed by ensemble musicians \shortcite{Glowinski2014,Keller2010,Ragert2013}; however, it is not always clear whether this coordination reflects interaction between musicians or simply exposure to the same musical stimulus. 

Some research has shown that audiences discriminate instances of motor-based interaction between ensemble members. For example, \shortciteA{Moran2015} found that observers could identify correctly-paired musicians based on motion cues during periods of solo improvisation (i.e., while one performer was playing and the other was ``back-channelling''). \shortciteA{Eerola2018} showed that coordination in periodic body movements is a salient cue to performer interaction that seems to affect audience perception. Furthermore, listeners have been shown to discriminate social intentions (e.g., dominance, insolence) in sound and motion recordings of improvised duo performances \cite{Aucouturier2017}.

Interaction can be quantified when musicians perform under assigned leader/follower conditions. \shortciteA{Chang2017} recorded body sway in members of a string quartet, and used Granger causalities to estimate the magnitude and direction of information flow between performers. Granger causality indicates how well one time series is predicted by a second time series, taking into account how well the first predicts itself \cite{Granger1969}. Larger Granger causalities indicate stronger prediction by the second time series, and a greater likelihood that the second time series influenced the first. The results of the study by \citeA{Chang2017} suggested that when (privately) assigned the role of leader, performers influenced their co-performers more and were influenced less than when following. \citeA{DAusilio2012} also used Granger causalities to assess potential patterns of influence between a conductor and ensemble of eight violinists. Conductor movements were found to modulate the strength of the interactions between violinists, showing how, in larger ensembles, interactions between performers form a complex network in which each person influences and is influenced by multiple others.

In smaller ensembles, leader/follower roles are not as clearly defined. Performers may take turns building on each other's ideas or contribute equally to shaping a new idea. Turn-taking behaviour in improvising duos was the focus of a study by \citeA{Walton2017}, which examined the number and strength of recurrent gestural patterns across performers' movement profiles. Cross-recurrence was higher -- indicating more interperformer coordination in movement -- when performers improvised over a drone backing track than when they improvised over a swing backing track, suggesting increased interaction during the more open-ended performance condition. 

Thus, interaction at the level of body movement can be tapped by analyses that identify instances of coordination in performed gestures, and analyses that estimate the magnitude of influence that one performer has over another, within a given window of time. The studies discussed above suggest that different performance conditions (e.g., genre, ensemble arrangement, performance instructions, piece structure) encourage different types and strengths of interaction. More broadly, performance conditions likely contribute to how much ensembles draw on signalling and interaction processes in order to produce cohesive, coordinated output \cite{MacRitchie2017}.

In the current study, we sought evidence of between-performer gestural interaction. Duo musicians' head movements were subjected to two categories of analysis. First, we examined how the gesture kinematics of individual performers varied across performances, types of piece structure, and visual contact conditions. It was hypothesized that when musicians are in visual contact with each other, they modify their gesture kinematics in ways that are likely to increase how meaningful, informative, and/or influential their movements are -- for example, by making movements of a larger magnitude or movements that are smoother. Second, we tested for instances of primo-to-secondo and secondo-to-primo influence and assessed coordination in head movements across the course of their performances. Interaction between performers was expected to heighten in certain parts of the performance and generally increase as performers rehearsed and became familiar with the music and their partner.

\subsection{Current study}
Much of the literature on ensemble performance has focused on the question of how musicians synchronize. Synchronization is not the sole requirement for a successful performance, however. While ensembles must at least entrain to a common metrical structure, precise synchronization of note onsets/offsets is sometimes sacrificed -- or even deliberately eschewed -- for the sake of expressivity (note that in the case of unmetered music, entrainment may still emerge unintentionally in the form of a complex hierarchy of periodic relationships; \citeNP{Clayton2007}).

The current study did not evaluate performance success; rather, we presented highly skilled musicians with a familiar task (to jointly rehearse a new piece in preparation for performance) and observed how they went about completing it. We assumed that our musicians would strive to perform well, allowing us to observe their behaviour as they worked towards and gave a successful performance (according to their own definitions of success). By presenting them with a custom-composed piece that was high in ``expressive ambiguity'' (see Methods), we forced them to have to make decisions about how to play (e.g., at what tempo, with which phrasing, etc.). The aim was to observe their behaviour as they went about making those decisions and negotiated a coherent interpretation of the piece with their partner.

The study investigated how signalling and interaction processes were affected by piece structure, rehearsal, and visual contact. Our hypotheses were as follows:

\begin{enumerate}
\item Signalling in the form of cueing gestures was expected to occur at moments of high temporal uncertainty (e.g., piece onset and following long held notes). As performers rehearsed the piece, it was hypothesized that they would integrate these cueing gestures into their performance. As a result, we expected to find increased consistency in gesture patterns during sections of the piece that were high in temporal irregularity.

\item We expected performers to show more movement and smoother movements during passages with irregular tempo. Quantity and smoothness of movement were also expected to increase with rehearsal. Between-performer coordination, likewise, was expected to increase in periods of temporal irregularity and with rehearsal.

\item Effects of visual contact were expected to manifest as differences in gesture kinematics between performances given under normal visual contact conditions and performances given under no-visual-contact conditions. Specifically, performers were expected to show increased quantity and smoothness of movement when they were aware that their partner might be watching. Such an effect would suggest an effort to interact and an effort to make themselves more predictable.

\item A number of factors influence the emergence of leader/follower roles during duo performance -- some are social (e.g., differences in experience, age, or personality) while others relate to the musical structure. In the current study, passages that were structured to comprise melody and accompaniment assignments were expected to encourage performers to assume leader/follower roles, and we hypothesized that these roles would be reflected in their body movements. Specifically, we hypothesized that performers would influence their partner more when leading than when following. Such a finding would be evidence that the mutual exchange of information that occurs during performer interaction can be biased in one direction or the other, causing the contributions of one performer to dominate. It would also be evidence that this bias can change directions from moment to moment, as playing conditions change.
\end{enumerate}

\begin{figure*}
\begin{center}
\includegraphics[width=\textwidth]{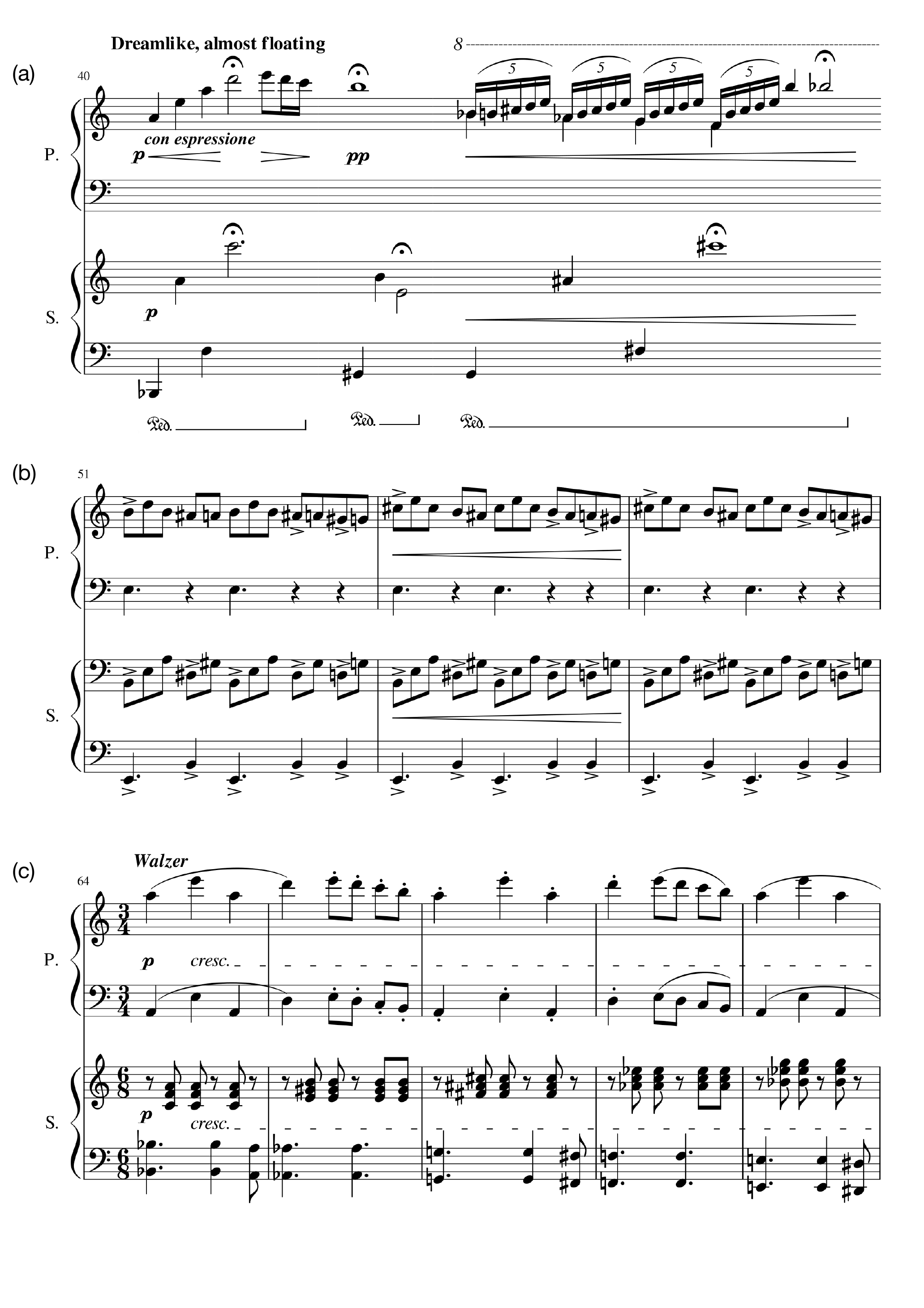}
\caption{Three excerpts from the piano version of the duet that performers played during the experiment. Line (a) is the start of the unmetered section, line (b) comes from a section with 5+7/8 meter where primo and secondo are intended to synchronize their accents, and line (c) is the start of the ending section, in which the main theme returns in the primo part.}
\label{fig:score}
\end{center}
\end{figure*}

\section{Methods} 

\subsection{Participants}
Ethical approval for this research was obtained from the University of Music and Performing Arts Vienna Ethics Committee. Forty-two professional musicians completed the experiment (20 pianists and 22 clarinettists). Pianists (age $M=28.3$ years; 14 female) reported an average of 21.6 years ($SD=7.3$) of formal training and 18.8 performances per year ($SD=16.7$). Clarinettists (age $M=24.0$ years; 13 female) reported an average of 15.0 years ($SD=4.2$) of formal training and 53.7 performances per year ($SD=53.2$). 

The high variability in the average numbers of performances per year is attributable to differences in the type of performances given and type/number of ensembles that musicians belonged to (e.g., some pianists performed regularly as accompanists while others had a few large-scale recitals per year). Clarinettists show especially high variability as a couple reported performing multiple times per week (e.g., in local orchestras as well as small ensembles). Many musicians already knew their partner prior to the experiment, but only some had played together previously (4 piano duos and 2 clarinet duos).

Participants provided written informed consent and received a small compensation. 

\subsection{Design}
This experiment tested the effects of three independent variables (piece structure; performance number; presence/absence of visual contact) on several aspects of duo performers' head movement. The piece structure variable tested the effects of temporal regularity and included four levels, corresponding to different sections of the piece (``entrance'', ``unmetered'', ``regular'', and ``ending'' sections; see sections 2.3 and 2.5.1 for descriptions). The performance number variable tested the effects of rehearsal and included three levels, corresponding to each of the three full recordings of the piece that were made before (1st), midway through (2nd), and after the rehearsal period (3rd) under normal visual contact conditions. The visual contact variable had two levels, corresponding to the 3rd and 4th performances, which were given under normal visual contact and blind conditions, respectively.

For between-performance analyses, the 3rd performance was treated as a ``reference'', since this was the final performance given under normal visual contact conditions, and thus expected to correspond most closely to the way the musicians ultimately intended the music to be played.

Several categories of dependent variables were investigated, corresponding to different aspects of performers' gesture kinematics. As within-performer measures, the 1) distribution of cueing gestures across performed beats, 2) consistency of gesture patterns across performances, and 3) quantity and 4) smoothness of gesture patterns were assessed. As within-duo measures, 5) periods of primo-to-secondo and secondo-to-primo influence and 6) coordination were assessed. All of these measures are defined in section 2.5.2.

\subsection{Stimuli and equipment}
Participants performed a piece composed by the second author specifically for this study. The piece was intended to be readily sight-read, but presented some challenges for duo coordination, including several metrical changes, unusual meters (e.g., 5+7/8), an unmetered passage, and accent patterns that the primo and secondo had to synchronize. The score included some expressive instructions (e.g., ``con espressione'', ``energico'', ``free''), but there were no explicit tempo markings. The piece was initially composed for two pianos, then arranged for two clarinets with input from a professional clarinettist. Some excerpts are given in Figure \ref{fig:score}, and the full piano score can be found in \citeA{Bishopsubmitted}.

\begin{figure*}
\begin{center}
\includegraphics[width=.7\textwidth]{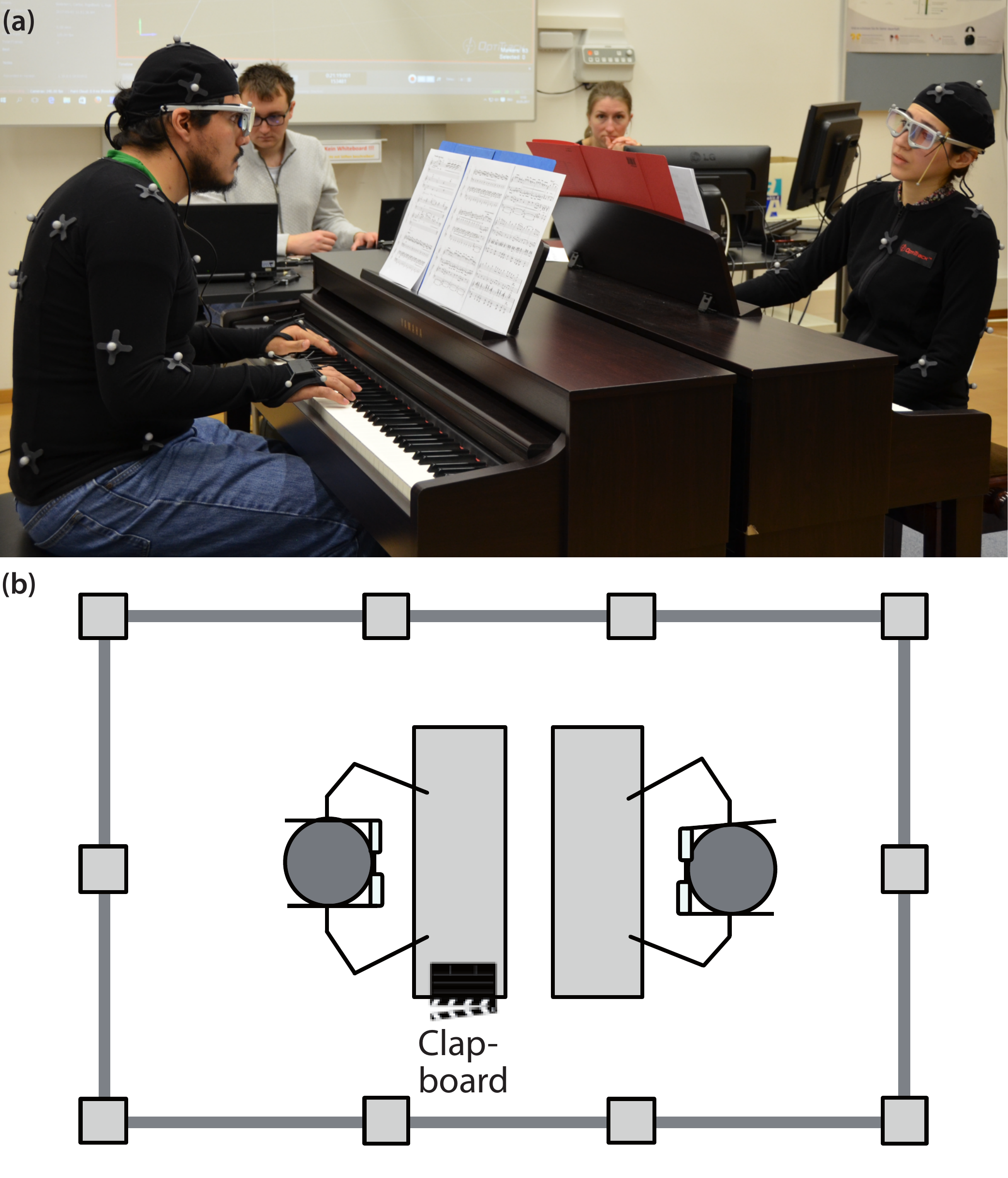}
\caption{Duo recording procedure. (a) A piano duo recording a performance and (b) a diagram of the recording set-up. Ten OptiTrack cameras lined the perimeter of the motion capture space, and duos sat (pianists) or stood (clarinettists) in the center. The clapboard was used to synchronize recorded audio with motion capture.}
\label{fig:setup}
\end{center}
\end{figure*}

A 10-camera (Prime 13) OptiTrack motion capture system, recording at 240 frames per second, tracked performers' upper body movements (Figure \ref{fig:setup}). Performers were fitted with 25 reflective markers, including 3 on the head. Clarinettists' instruments were also fitted with 4 markers, two near the mouthpiece and two near the bell. Eye gaze data were simultaneously collected using SMI ETC 2 wireless glasses (gaze data are reported in \citeNP{Bishopsubmitted}).

Pianists performed on Yamaha Clavinovas, and audio/MIDI data were collected using a Focusrite Scarlett 18i8 sound card and recorded as separate tracks in Ableton Live. Clarinettists performed on their own instruments. Their audio was collected using DPA d:vote 4099 clip-on microphones and likewise recorded as separate tracks. To synchronize audio with OptiTrack recordings, a film clapboard was marked with reflective markers and placed in view of the cameras, near to a microphone that collected audio from the room. The clapboard was struck once at the start and end of each recording, and all recordings were trimmed retrospectively to between these points.

\subsection{Procedure}
Performers were presented with musical scores at the start of the session, randomly assigned to primo and secondo parts, and instructed to practice together in preparation for recording some polished performances. They were not asked to memorize the music. We explained that our aim was to investigate performer interaction during rehearsal of an unfamiliar piece.

Performers were positioned facing each other, about 1.5 meters apart. Clarinettists played standing and thereby had somewhat more freedom of movement than did pianists. An initial sight-read performance of the piece was recorded first, for which duos were encouraged to continue playing regardless of errors. This performance was followed by 10--20 minutes of free duo rehearsal (the decision to proceed was made jointly by the performers and the experimenter, when the performers seemed to be about halfway to achieving a polished performance; some duos needed more time than others). A second performance was then recorded, followed by 10--20 more minutes of rehearsal\footnote{Practice time did not correlate with any of our dependent variables.}. Two polished performances were recorded at the end of the session -- one with visual contact between performers, then one without. Finally, all participants completed a short questionnaire on their musical background and perceptions of the experiment. Each recording session took about one hour altogether.

\subsection{Analysis}

\subsubsection{Note onsets}
MIDI data from piano performances were matched to the score using a score-performance matcher that pairs performed pitches with score notes based on pitch sequence information \shortcite{Flossmann2010}. Output from the matcher lists the onset times for correctly performed notes. For clarinet recordings, note onsets at section boundaries (see below) were identified manually using waveform and spectrogram information. Beat positions within sections were then interpolated. Note onset times were aligned with motion capture data by trimming the start of the recordings, so that they began at the moment that the clapboard was struck (see Stimuli and equipment).

\begin{figure*}
\begin{center}
\includegraphics[width=.9\textwidth]{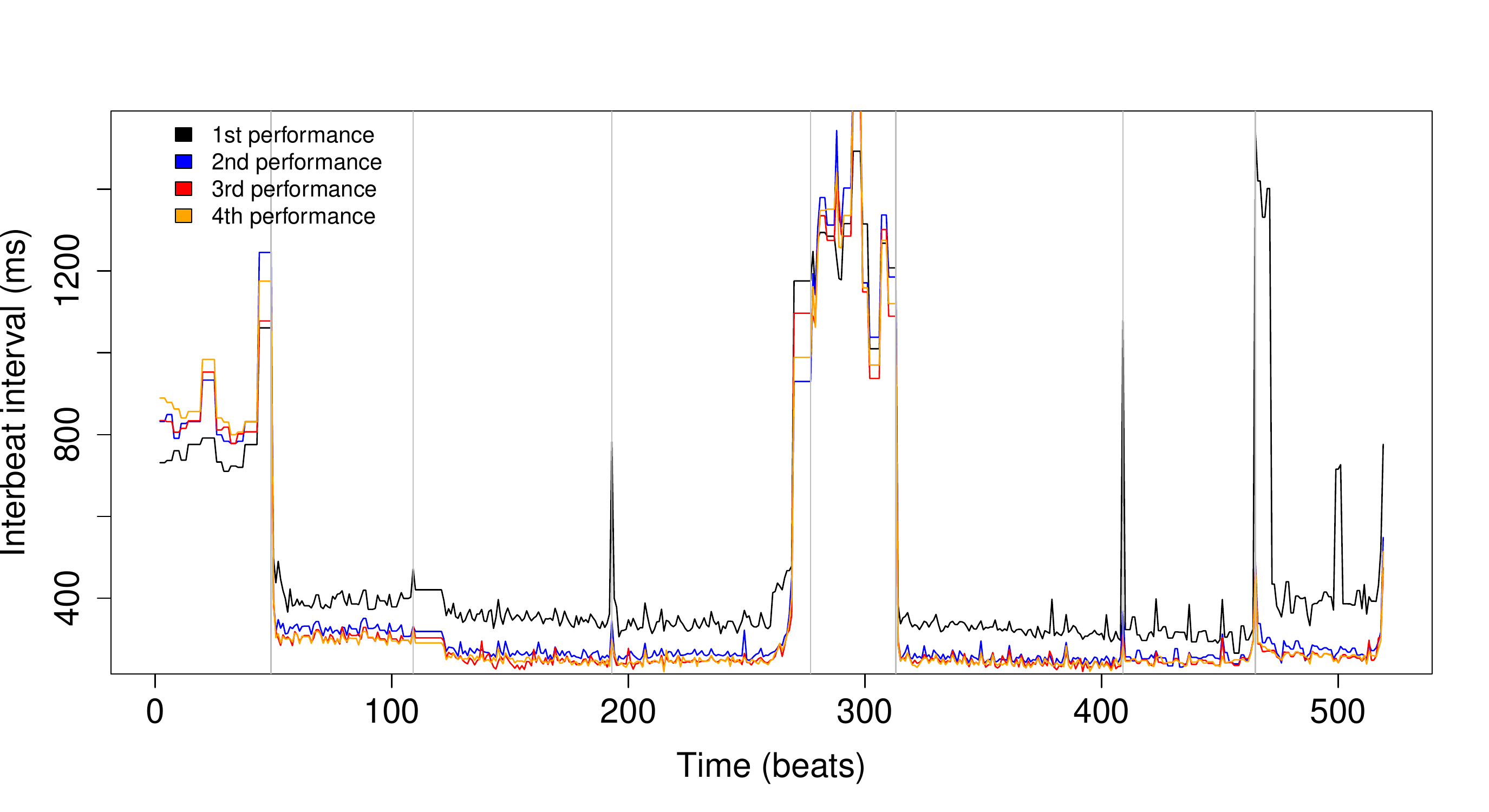}
\caption{Average tempo curves for pianists' performances. Interbeat intervals were computed for each duo using only the notes for which both primo and secondo onset times were available. Primo and secondo note onset times were then averaged, and a full series of interbeat intervals was computed using linear interpolation.  Interbeat intervals were then averaged across duos at each timestep to achieve the curves presented in the figure. Vertical grey bars indicate metrical changes. Beats 1--49 comprise the entrance section, beats 277--313 the unmetered section, and 465--519 the ending; the rest of the piece was regularly timed.}
\label{fig:timing_curve}
\end{center}
\end{figure*}

The piece was divided into sections, with boundaries placed wherever the meter changed. The first 8 bars (``entrance''), the final 10 bars (``ending''), and the unmetered section (``unmetered'') were shown by \citeA{Bishopsubmitted} to be less temporally stable than the more regularly-timed (``regular'') sections. Temporal stability, in that study, was assessed in terms of within-performance tempo variability and the magnitude of primo-secondo asynchronies. Differences in tempo between piece sections can be seen in Figure \ref{fig:timing_curve}, which shows tempo curves for pianists' performances, averaged across duos for each of performances 1--4.

\subsubsection{Body movement}
We restricted our analysis to musicians' head movements, as these were the only source of visual motion cues that were available to all duos (clarinettists could see their partner's arms and torso to some degree, but pianists could only see their partner's head and shoulders). Position data were smoothed using functional data analysis \cite{Goebl2008,Ramsay2005}. Order-6 b-splines were fit to marker trajectories with knots every 50 ms and a roughness penalty applied to the fourth derivative, which smoothed the second derivative (acceleration). Functional data were then converted back to position, velocity, and acceleration series, with samples every 5 ms.

Several measures were extracted from these resampled gesture data for further analysis. These measures are described below (see Appendix for more technical detail), along with the linear mixed effects modelling procedure that we used to evaluate effects of piece section and rehearsal.

\textbf{Cueing gestures.} Performers' \emph{head acceleration} curves were examined to determine whether cueing gestures were given during the performance. We focused on acceleration (rather than velocity or trajectory), since acceleration patterns have previously been shown to communicate beat position in gestures given by conductors \cite{Luck2008a,Luck2009} and instrumentalists \cite{Bishop2017,Bishop2018}.

We began by identifying performances in which the performer gave a ``cueing-in'' gesture in the form of a head nod at piece onset. Both primo and secondo performances were considered to maximize the number of performances that could be analyzed (we noticed during data collection that cueing-in gestures were usually given by the primo, but sometimes given by the secondo).

Using dynamic time warping, within-performance alignments were then made between the acceleration pattern of the cueing-in gesture and the acceleration pattern occurring at each subsequent (quarter note) ``test'' beat throughout the remainder of the performance. This analysis yielded a distance value for each test beat that represented its similarity to the beat containing the cueing-in gesture (``test-reference distance''). Small test-reference distances indicated greater similarity, and a high likelihood that the test segment contained a cueing gesture similar to that given at piece onset. (See Appendix for further detail on cueing-in gestures identification and dynamic time warping.

We hypothesized that the cueing gestures would appear at particular points in the performances. A total of eight key locations in the score were identified (in addition to the moment before piece onset), corresponding to the moment before the start of the unmetered section and a further 7 points in the unmetered section where held notes (marked with a fermata) occurred.

To test whether cueing gestures occurred reliably at these locations, we made an analysis to compare the test-reference distances for these 8 quarter note beats with random samples of 8 quarter note beats taken from each of the entrance, regular, and ending sections. This random sampling method allowed us to compare samples of equal size. Our prediction was that test-reference distances would be lower for the key points in the unmetered section (where cueing gestures should occur) than for the samples of points taken randomly from other sections (where no cueing gestures were expected). The effects of piece section and rehearsal on test-reference distances were tested using linear mixed effects modelling (see below).

\textbf{Consistency} of \emph{gesture trajectories} across performances was calculated using dynamic time warping. Our focus was on gesture trajectory for this analysis (rather than velocity or acceleration) because we wanted to know whether performers would establish a gesture script and reiterate very similar movement patterns across successive performances (whereas similar patterns of velocity or acceleration could be achieved via different gesture trajectories).

The first, second, and fourth performances were taken as ``test'' performances and aligned with the third ``reference'' performance (see section 2.2). Alignments were made per section, between performances (e.g., the second performance ending was aligned with the third performance ending). Distances between the aligned test and reference sections (``test-reference distances'') served as our measure of consistency. (See Appendix for further detail on the dynamic time warping procedure.)

\textbf{Quantity of motion} was calculated as the mean absolute distance that the head travelled (in m) per 5 ms observation within each performance section. We chose to use mean (rather than summed) distance values so that quantity of motion could be compared between piece sections that differed in length. (See Appendix for a mathematical definition.)

\textbf{Smoothness} was operationalized in terms of jerk \cite{Woellner2012}. The first derivative of each series of accelerations was computed. The resulting series of values, referred to as jerk, indicates change in acceleration. High jerk -- or rapid changes in acceleration -- corresponds to low smoothness.

\textbf{Direction of influence} was estimated using Granger Causalities (``g-causalities''), which were calculated for \emph{head acceleration} curves across each performance, using a rolling window method. G-causalities provide an estimation of the influence that one time series has over another, and has previously been used to explore leader/follower relationships in music ensembles \cite{Chang2017,DAusilio2012}. We chose to test for influence at the level of head acceleration (rather than velocity or trajectory) given the relationship between acceleration and musical timing \cite{Bishop2017,DAusilio2012,Himberg2011}.

In the current study, g-causality was measured in both directions for each window; that is, we measured the potential influence of the primo on the secondo and the potential influence of the secondo on the primo. Possible outcomes therefore included a significant result in one direction or the other (indicating a high likelihood that one performer influenced the other), no significant results, or significant results in both directions. Cases where significant results were achieved in both directions were treated as non-significant, as these cases tend to occur when there is an external factor simultaneously influencing both variables. (Further detail on our procedure is given in the Appendix.)

The results of the g-causality tests were used to test the hypothesis that leader-to-follower influence would outweigh follower-to-leader influence in sections of the piece where leader/follower relationships were implied by a melody/accompaniment structure. For each of these sections, three measures were calculated: 1) the proportion of beats where only the primo influence on the secondo was significant, 2) the proportion of beats where only the secondo influence on the primo was significant, and 3) the proportion of beats where both or neither direction of influence was significant. These calculations were made separately for each performance. The dependent variable for this analysis was the proportion of beats for which a significant result was obtained.

\textbf{Between-performer coordination.} Cross-correlations were computed between performers' \emph{head acceleration} curves as a measure of between-performer coordination. We focused on head acceleration, again, because of its relationship with musical timing. Cross-correlations were computed using a rolling window method across pairs of performances, and were scaled to account for performers' quantity of movement (otherwise windows containing very little movement would result in seemingly high cross-correlations; see Appendix for further explanation). Absolute scaled lag 0 cross-correlation values were then averaged per piece section.

\textbf{Linear mixed effects modelling (LME).} We used LME (``nlme'' package in R) to test the effects of piece section and rehearsal on the measures described above. To account for repeated measures, piece section nested within performance number nested within subject was included in each model as a random effect. Only data for performances 1--3 were included in these analyses; comparisons between the third and fourth performances were made separately to test for effects of visual contact. Where post-hoc tests were made (using the ``lsmeans'' package in R), Bonferroni corrections were applied to the significance level: tests of piece section were evaluated at $\alpha=.02$ and tests of rehearsal were evaluated at $\alpha=.008$.

\section{Results}
The results presented in this section are organized into three sub-sections. The first subsection investigates the distribution of cueing gestures across performances and tests how consistent gesture patterns were across performances. The second subsection approaches interactive behaviour from a within-performer perspective and investigates how gesture quantity and smoothness varied according to piece structure and performance number. The third section examines within-duo interaction by evaluating movement coordination across piece sections and performances and examining how performers influenced each other.

\subsection{Location and consistency of cueing gestures}
\subsubsection{Location of cueing gestures}
\begin{figure*}
\begin{center}
\includegraphics[width=.9\textwidth]{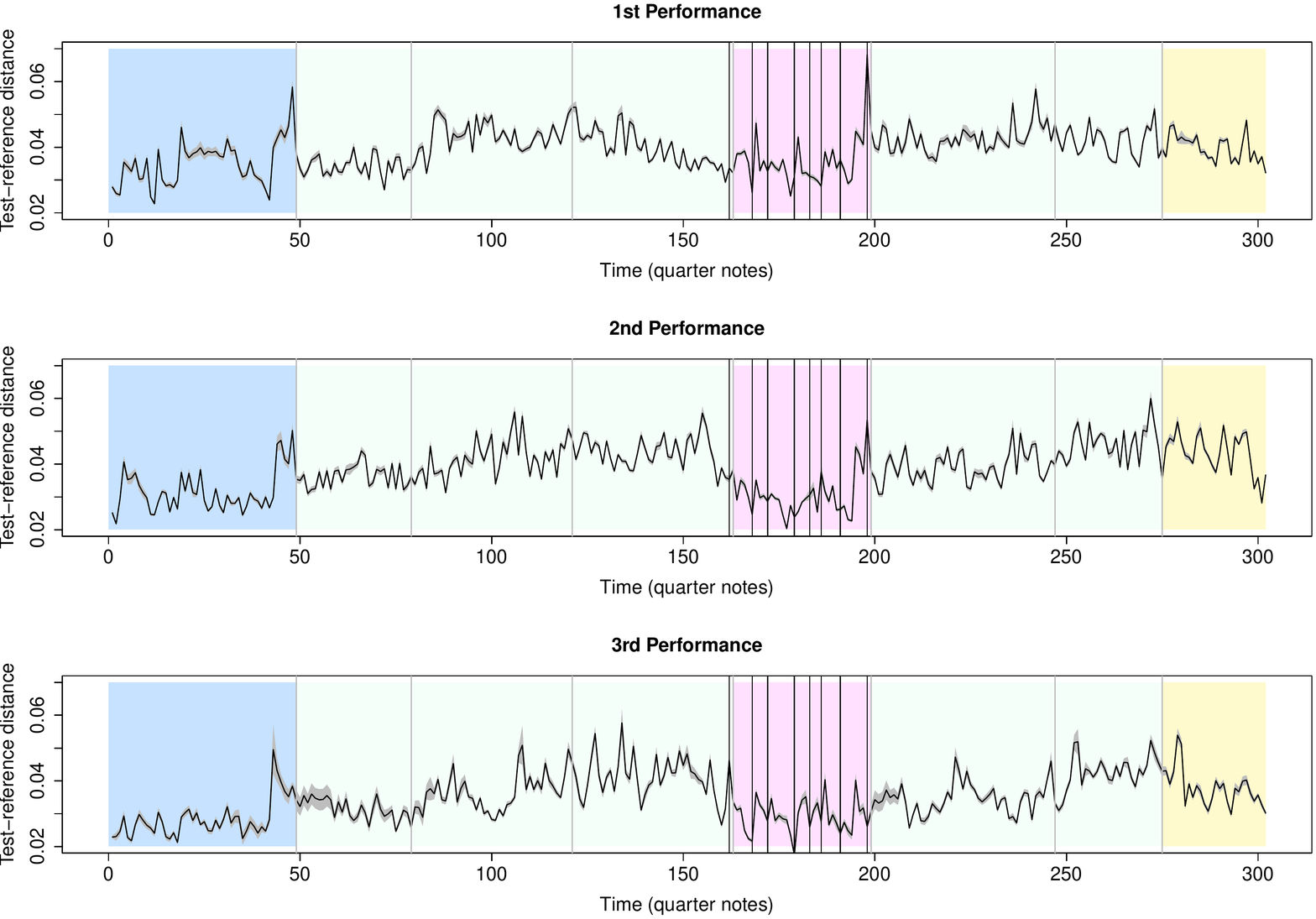}
\caption{Locations of cueing gestures. Plots show mean normalized distances (computed within performers and averaged across performers), indicating the similarity between per-quarter note acceleration curves and the acceleration curves of performers' cueing-in gestures. Low normalized distances correspond to high similarity. The grey area behind the curve indicates variance. Piece sections are colour coded with blue indicating the entrance, green the regular sections, pink the unmetered section, and yellow the ending. Grey vertical lines indicate section boundaries and black vertical lines indicate the position of expected cueing gestures. }
\label{fig:dtw_cueing}
\end{center}
\end{figure*}

It was hypothesized that performers would give cueing gestures at specific points (corresponding to pauses) in the unmetered section of the piece. Figure \ref{fig:dtw_cueing} shows the average per quarter note test-reference distances that were computed for the first three performances (i.e., those given under normal visual contact conditions), as described in section 2.5.2. LME tested the effects of piece section and performance number on test-reference distances.

This test yielded a significant effect of piece section, $F(3,285)=9.84,p<.001$. The effect of performance number and the interaction were both nonsignificant ($p>.05$). Post-hoc tests indicated that distances were lower in the unmetered section than in the regular, $t(285)=3.35,p=.005,\eta^2=.04$, and ending sections, $t(285)=4.86,p<.001,\eta^2=.08$. Distances were similar in the unmetered and entrance sections, $t(285)=.71,p=.89$.

Thus, test-reference distances tended to be smaller at the points in the unmetered section that were identified as likely locations of cueing gestures than at randomly-selected points in regular and ending sections, as hypothesized. Test-reference distances were similar at the selected points in the unmetered section and randomly selected points in the entrance section, suggesting that cueing gestures may have been exchanged during the entrance section as well. 

\newpage

\subsubsection{Consistency of gesture patterns across performances}
It was hypothesized that duos would incorporate some visual cues in the form of body gestures into their performance plan and produce these gestures consistently across performances. As a result, increasing consistency in motion patterns was expected across performances (e.g., the second and third performances should be more similar than the first and third). These planned cues were expected to occur primarily in the entrance and unmetered sections of the piece, where visual signalling was thought to be more important.

\begin{figure}
	\subfloat[Consistency. Test-reference distances between head trajectories in the ``reference'' third performance and each of the first, second, and fourth performances. Error bars indicate standard error.]{{\includegraphics[width=.5\textwidth]{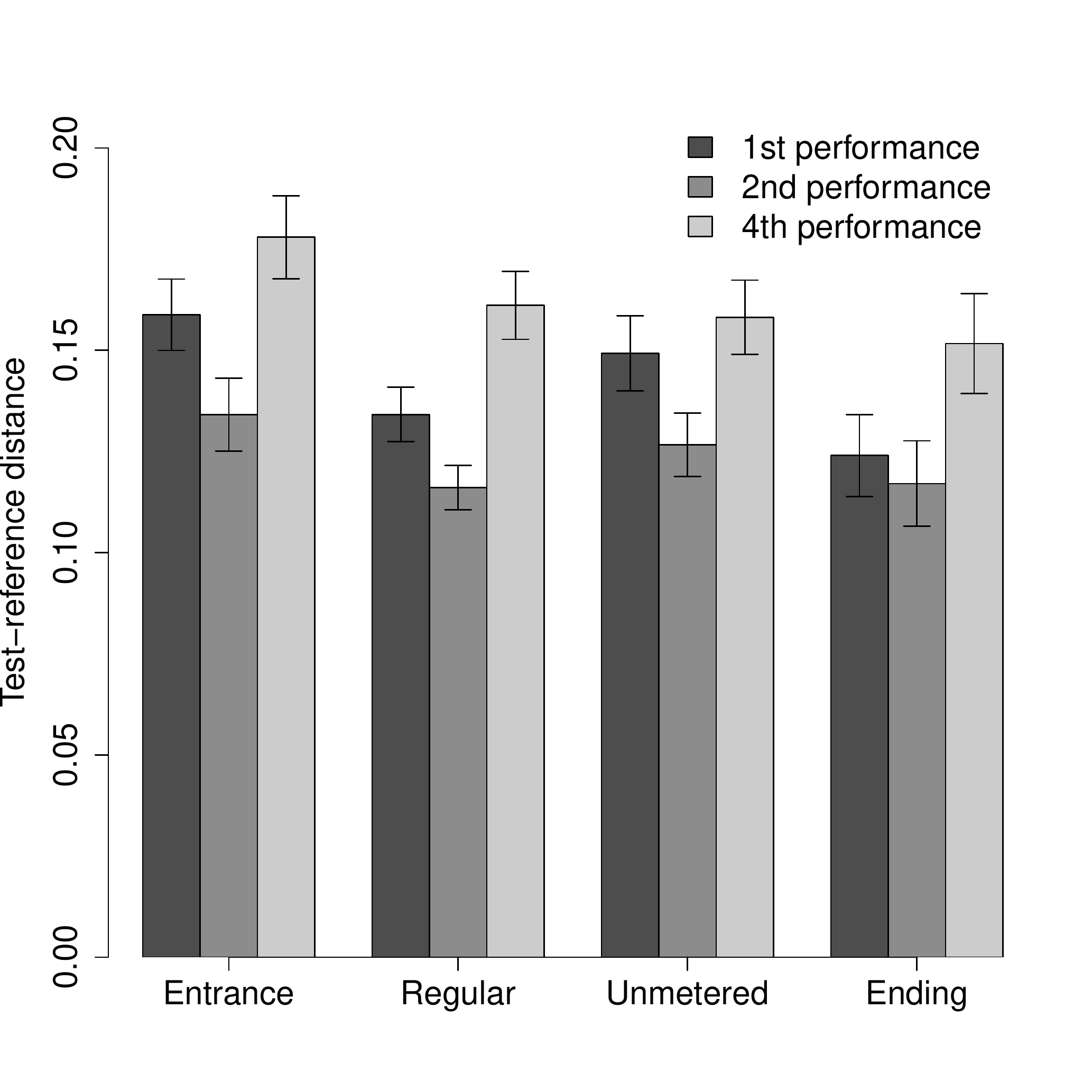}}
	\label{fig:consistency}}%
	\qquad
	\subfloat[Quantity of motion. Mean distances per observation travelled by performers' heads across piece sections and performances. Error bars indicate standard error.]{{\includegraphics[width=.5\textwidth]{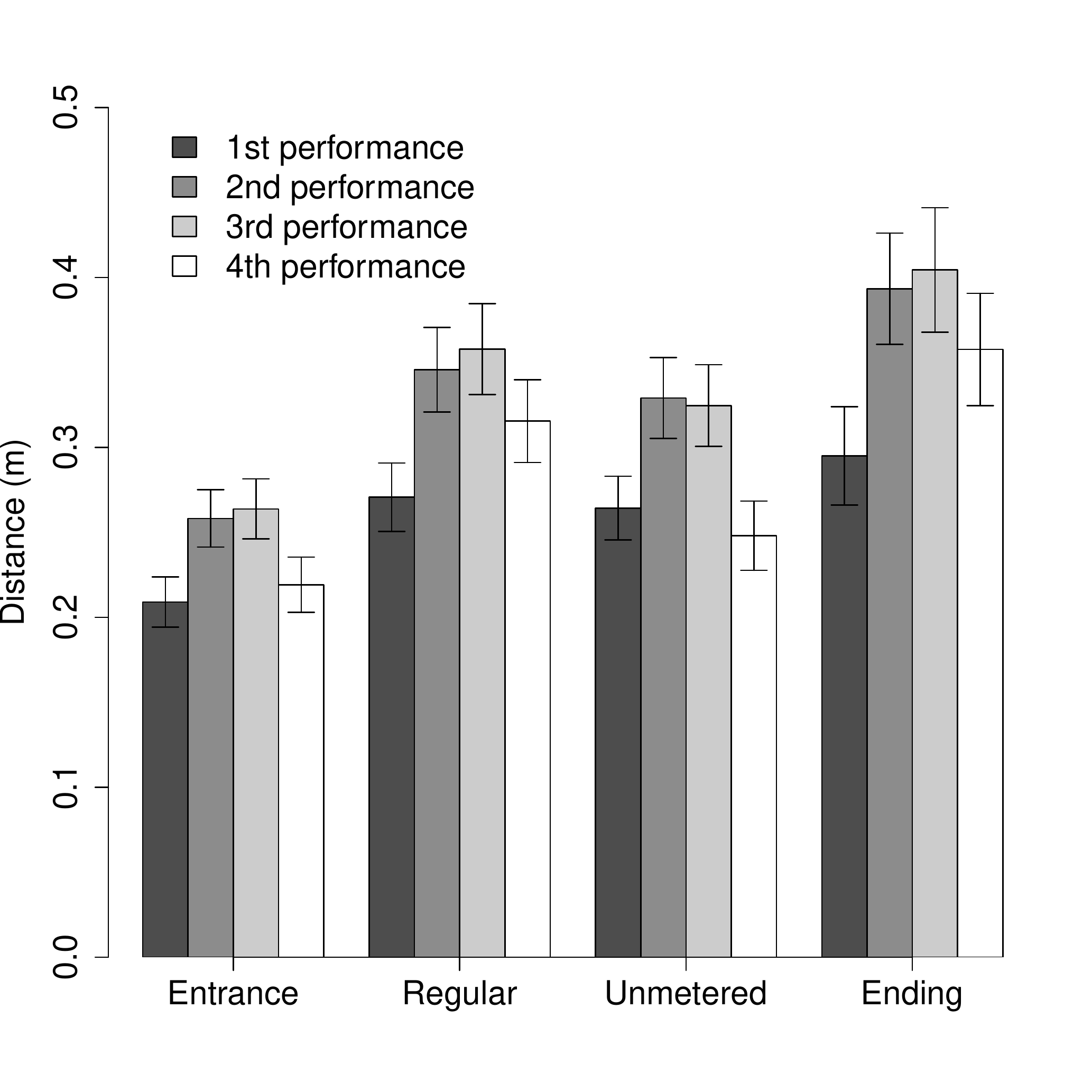}}
	\label{fig:qofm}}%
	\qquad
	\subfloat[Smoothness. Median absolute smoothness values for head motion across piece sections and performances. Smoother movements correspond to low values and jerky movements correspond to high values. Error bars indicate standard error.]{{\includegraphics[width=.5\textwidth]{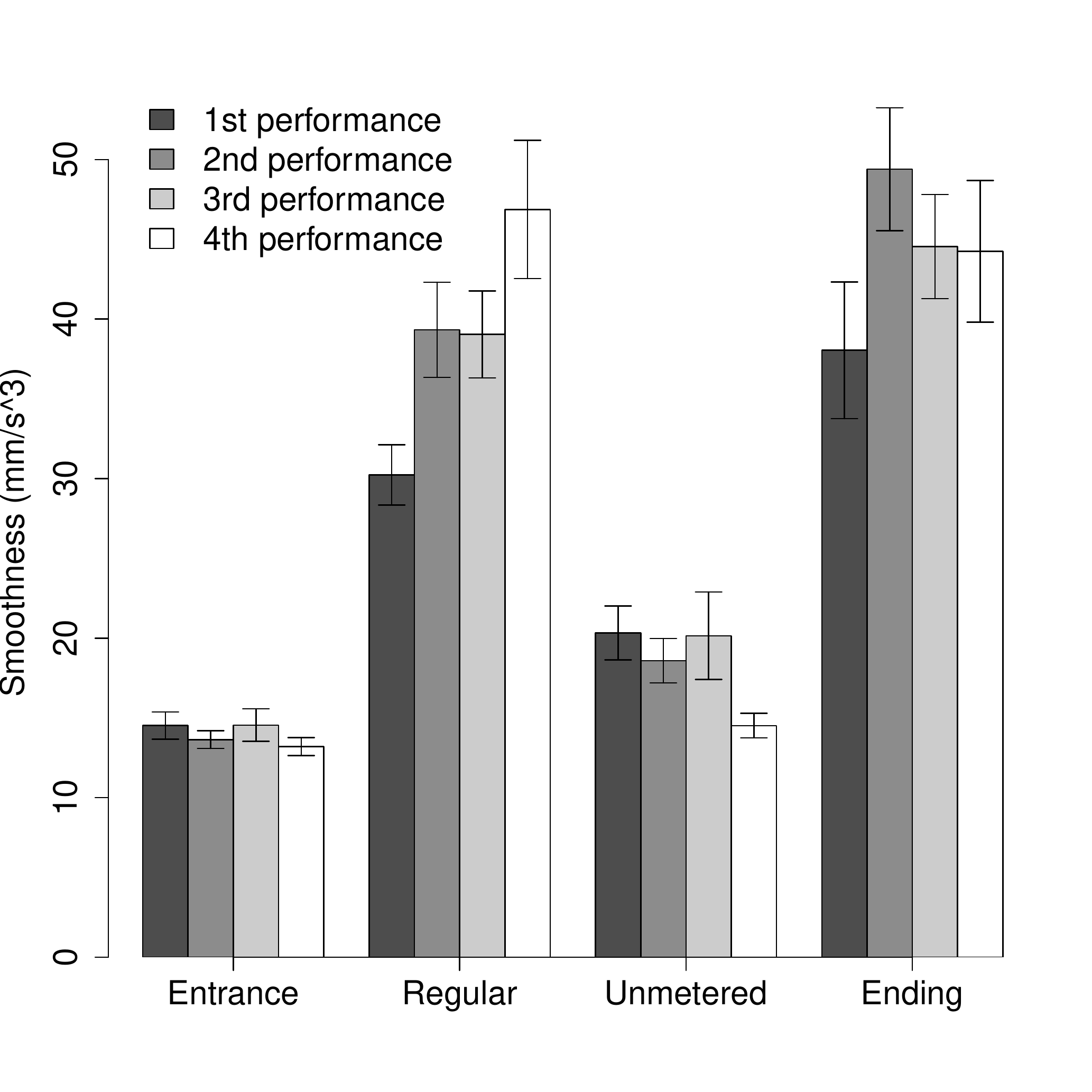}}
	\label{fig:smoothness}}%
	\qquad
\caption{Results of within-performer analyses. Note that while graphs include data for all performances, the 4th performance was excluded from analyses of rehearsal effects. The 3rd and 4th performances were compared separately, where appropriate, to test for effects of visual contact.}
\end{figure}

Figure \ref{fig:consistency} shows average test-reference distances by piece section and test performance. LME indicated significant effects of piece section, $F(3,233)=4.68,p=.003$, and performance number, $F(1,39)=16.40,p<.001$, but no interaction ($p>.05$).

The effect of performance number meant that test-reference distances were significantly smaller during the second performance than during the first. Post-hoc tests investigated the effect of piece section, but none of these yielded significant results. An additional t-test evaluating the effect of visual contact showed that the fourth performance differed more from the reference than did the second performance, $t(159)=6.5,p<.001,\eta^2=.21$. 

These results do not support our hypothesis that movements during temporally unstable passages are more consistent than movements during regularly timed passages. On the other hand, the results do support our hypothesis that performers' body movements grow more consistent as they rehearse: gesture trajectories during the second performance were more similar to the third performance than were gesture trajectories during the first performance. Also as expected, performers' movement patterns seemed to change once they could no longer see each other: the fourth performance was less similar to the third performance than the second was.

\subsection{Within-performer variations in movement features}

\subsubsection{Quantity of motion}
We hypothesized that quantity of motion would be higher during piece sections with irregular timing than during sections with regular timing, and higher after rehearsal than before. Figure \ref{fig:qofm} presents mean quantity of motion values by piece section and performance. LME indicated significant effects of piece structure, $F(3,333)=30.15,p<.001$, and performance number, $F(2,72)=40.67,p<.001$, but no interaction ($p>.05$).

Post-hoc tests showed that performers moved a greater distance in regular, $t(333)=7.8,p<.001,\eta^2=.15$, unmetered, $t(333)=4.7,p<.001,\eta^2=.06$, and ending sections, $t(333)=9.1,p<.001,\eta^2=.20$, than in the entrance section. They also moved more in the ending section than in either the regular, $t(333)=3.8,p=.001,\eta^2=.04$, or unmetered sections, $t(333)=4.4,p<.001,\eta^2=.05$. The regular and unmetered sections did not differ. 

Performers also moved a greater distance in the second, $t(72)=6.0,p<.001,\eta^2=.10$, and third performances, $t(72)=6.7,p<.001,\eta^2=.12$, than in the first performance. The second and third performances did not differ. 

A t-test showed that distances were greater in the third performance than in the fourth, $t(143)=5.2,p<.001,\eta^2=.16$, suggesting an effect of visual contact.

A follow-up test was made to check whether between-section or -performance differences in quantity of movement could be accounted for by differences in tempo (e.g., a fast tempo might encourage more movement than a slow tempo). LME was run with piece section, performance number (1--3), and tempo (mean interbeat interval duration) as fixed effects. Repeated measures were accounted for as described in section 2.5.2. This analysis yielded significant effects of piece section, $F(3,171)=14.65,p<.001$, performance number, $F(2,38)=21.66,p<.001$, and tempo, $F(1, 288)=10.55,p=.001$, but no interactions (all $p>.05$). Thus, while tempo affected quantity of movement, it could not explain differences between piece sections or performances.

\subsubsection{Smoothness of motion}
Smoothness was hypothesized to increase during piece sections with irregular timing and following rehearsal. Figure \ref{fig:smoothness} shows median absolute smoothness values by piece section and performance. LME indicated significant effects of piece section, $F(3,333)=77.85,p<.001$, and performance number, $F(2,72)=9.70,p<.001$, but no interaction ($p>.05$).

Post-hoc tests showed that motion was smoother in the entrance section than in regular, $t(333)=12.4,p<.001,\eta^2=.32$, unmetered, $t(333)=4.1,p<.001,\eta^2=.05$, and ending sections, $t(333)=10.1,p<.001,\eta^2=.24$. Motion was also smoother in the unmetered section than during regular, $t(333)=11.5,p<.001,\eta^2=.28$, and ending sections, $t(333)=8.6,p<.001,\eta^2=.18$. Regular and ending sections did not differ. None of the comparisons for performance number reached significance.

A t-test showed no difference in smoothness between the third and fourth performances, $t(143)=.7,p=.50$, indicating no effect of visual contact.

These findings are consistent with our hypothesis that movement is smoother in passages where visual interaction was more likely -- specifically, the entrance and unmetered sections -- than in regular sections.

\subsection{Within-duo patterns of influence and coordination}

\subsubsection{Leadership and direction of influence}

We hypothesized that the leader/follower roles that were implied by melody/accompaniment assignments in the score would be reflected in the way performers influenced each other's body movements. Specifically, performers were expected to influence their partners' movements more when playing the melody (leading) than when playing the accompaniment (following). 

LME tested for an interaction between melody/accompaniment section (primo playing the melody; secondo playing the melody) and direction of influence (primo influencing secondo; secondo influencing primo), and an effect of performance number. Performance number nested within duos was included as a random effect. This analysis yielded no significant effects ($p>.05$). Performers were therefore no more likely to influence their partners' head movements when playing the melody than when playing the accompaniment.

LME was also run to test for effects of piece structure and performance number. This analysis yielded no significant effects ($p>.05$).

\begin{landscape}
\begin{figure}
\centering
\vspace{-1cm}
	\subfloat[First performance.]{{\includegraphics[width=.48\linewidth]{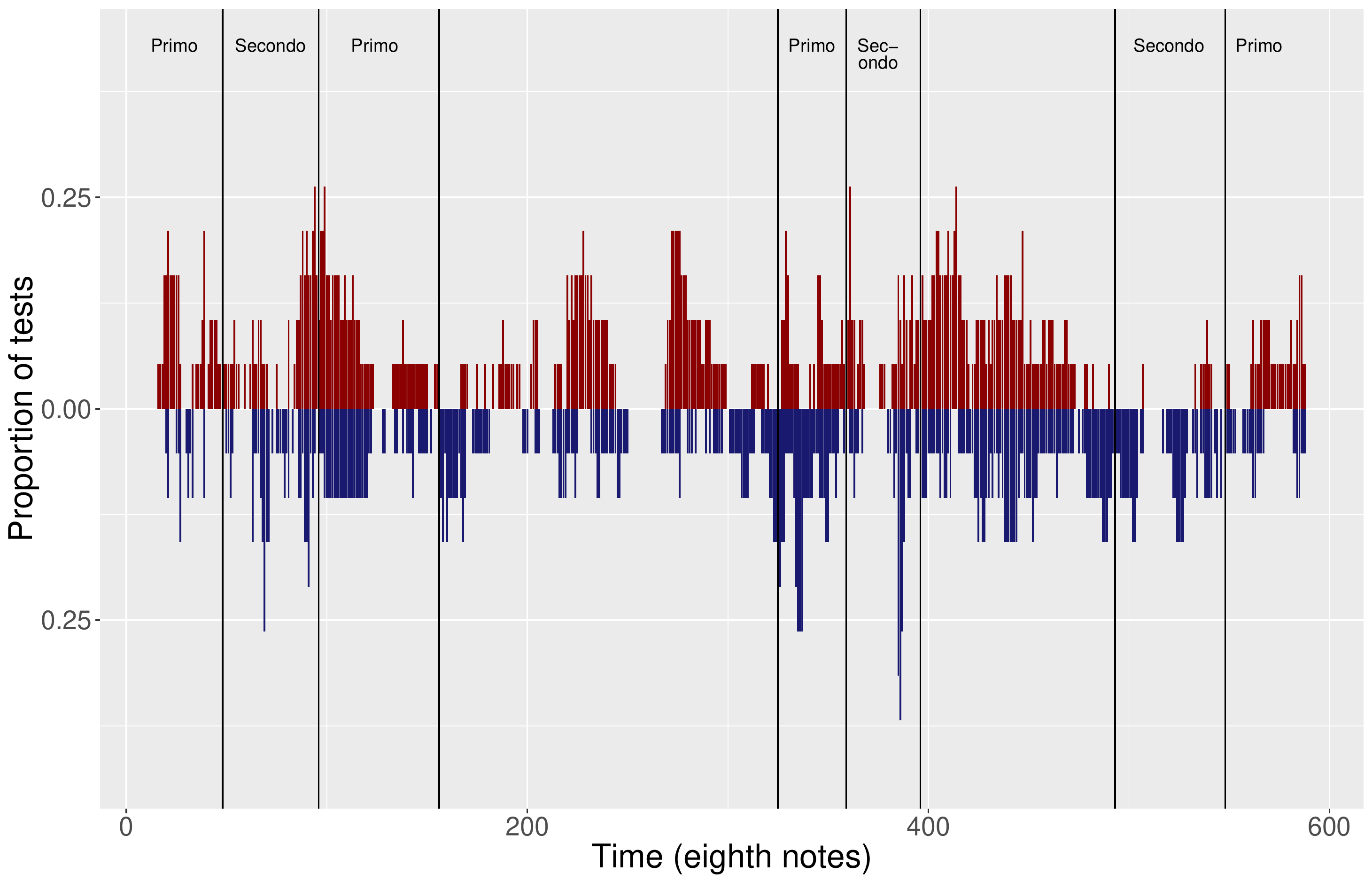}}}%
	\qquad
	\subfloat[Second performance.]{{\includegraphics[width=.48\linewidth]{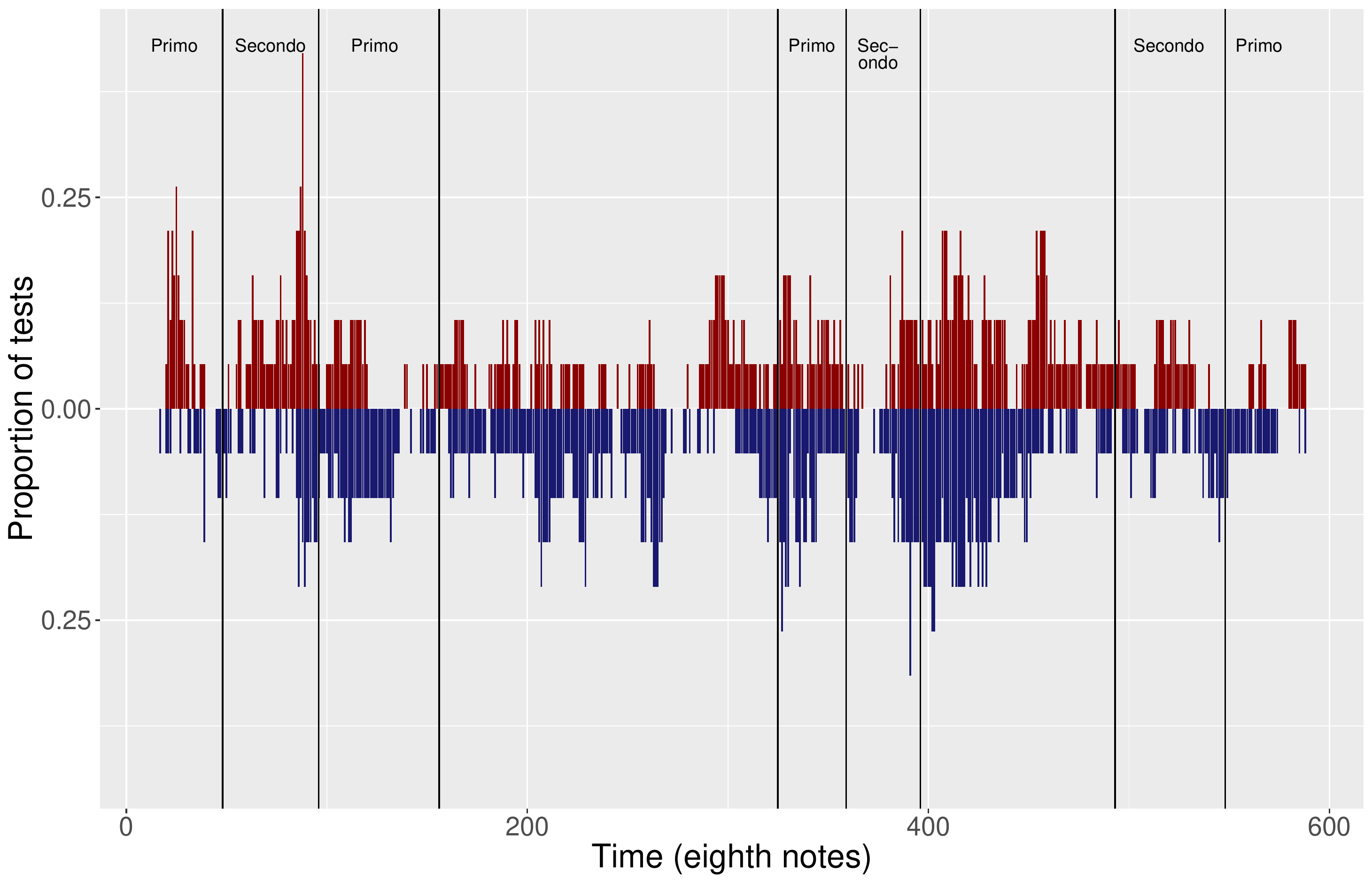}}}%
	\qquad
	
	\vspace{-.4cm}
	\subfloat[Third performance.]{{\includegraphics[width=.48\linewidth]{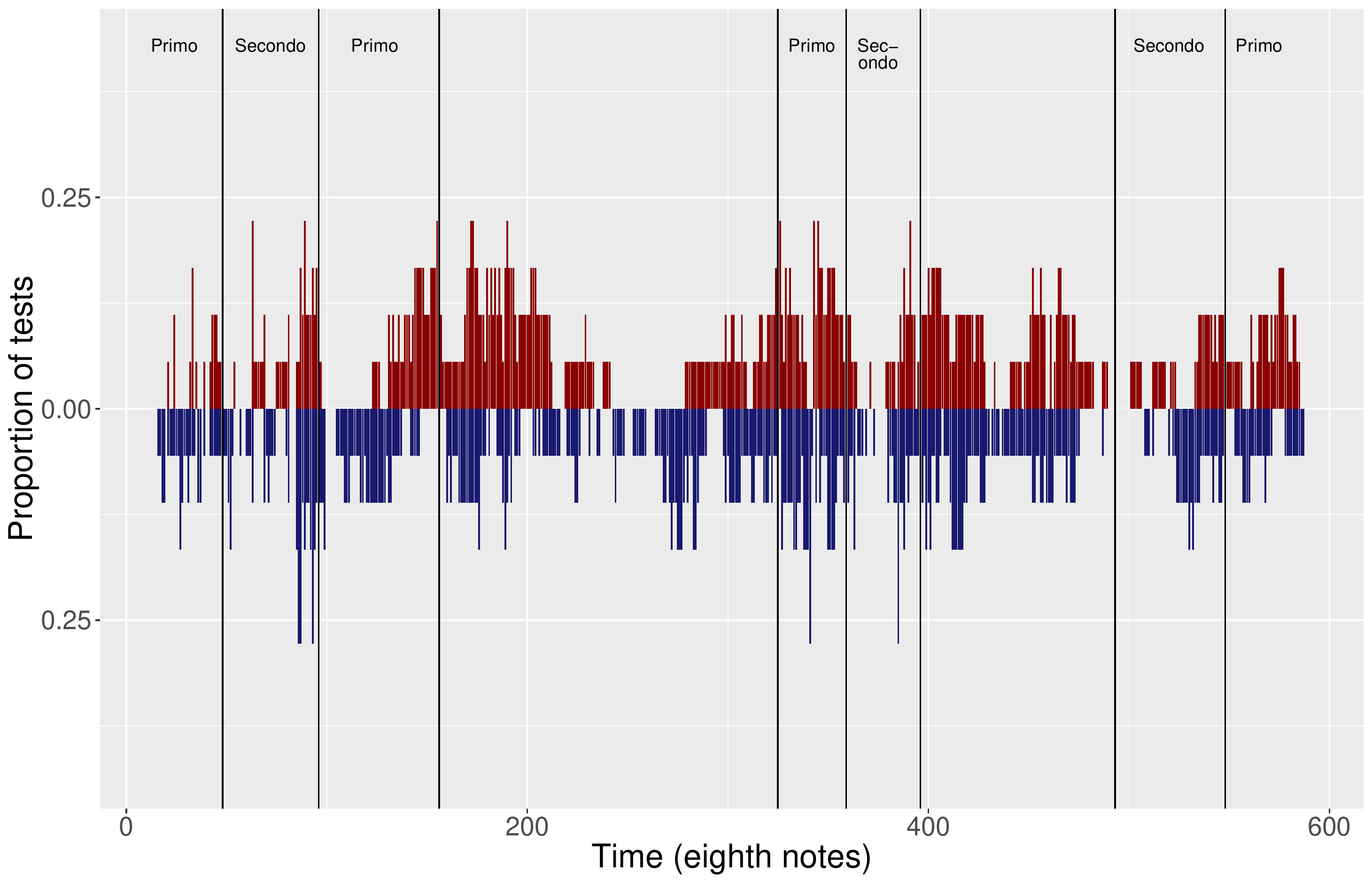}}}%
	\qquad
	\subfloat[Fourth performance.]{{\includegraphics[width=.48\linewidth]{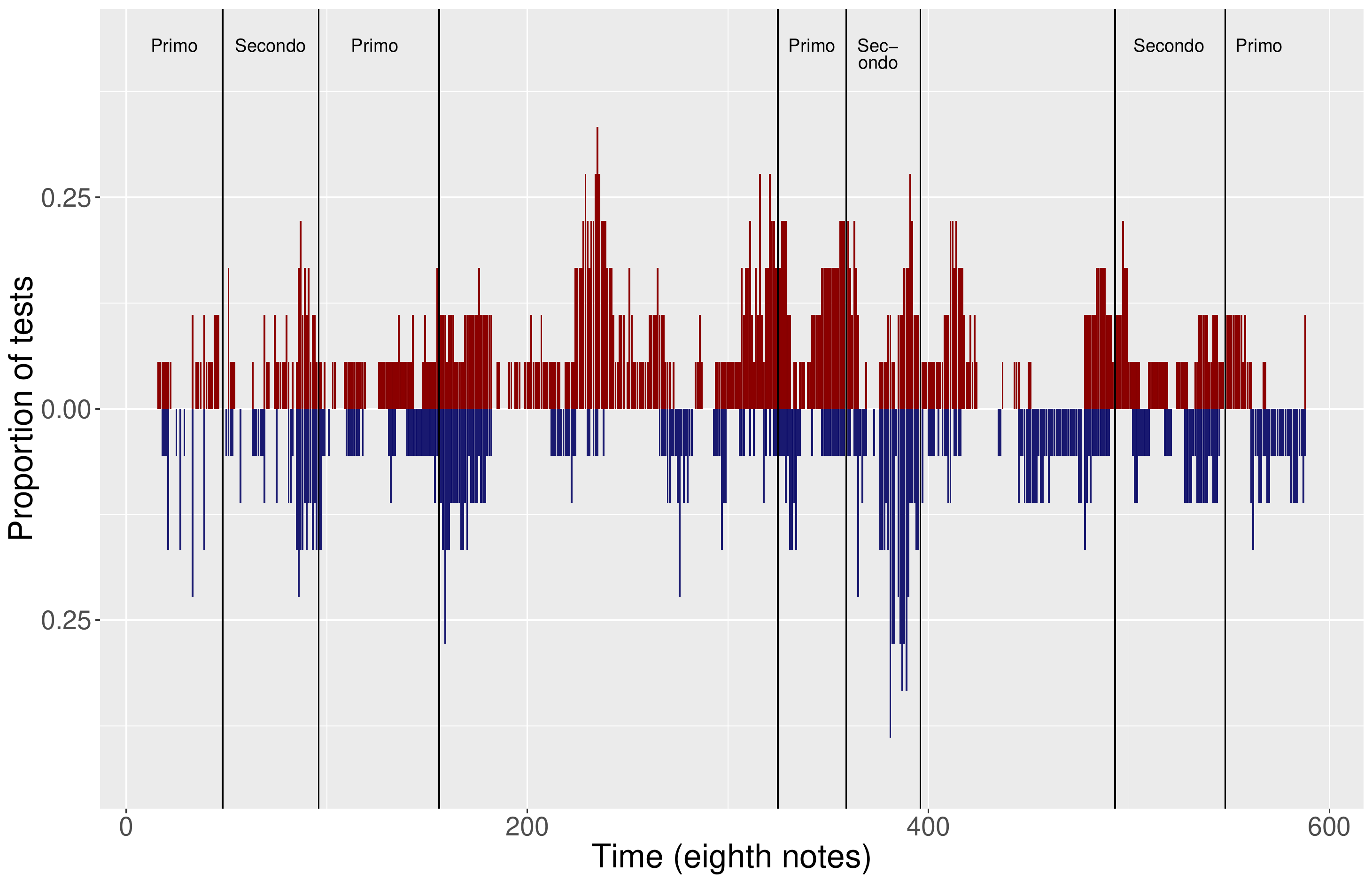}}}%
	\qquad
	\vspace{-.3cm}
\caption{Leadership and direction of influence. The plots present the proportion of g-causality tests per beat for head acceleration, averaged across duos, for which a significant result was achieved indicating primos influencing secondos (pointing up, red) or secondos influencing primos (pointing down, blue). X-axis values correspond to the middle beat of each (30 lag) window. Passages in which either the primo or secondo played the melody are marked with vertical black lines and labelled.} \label{fig:granger}
\end{figure}
\end{landscape}

Figure \ref{fig:granger} shows how the likelihood of performers influencing each other changed over the course of the four performances. The values presented here were obtained by running g-causality tests over a rolling window of 30 beats across each performance, using head acceleration data. Models contained one predictive lag. Passages of the piece with primo or secondo melody assignments are marked on the plots. The first primo- and secondo-lead passages listed on the plots together made up the entrance section, and the primo- and secondo-lead passages between beats 300--400 made up the unmetered section. The final primo-lead passage comprised the ending section, and the remaining passages were regular.

We report above a lack of interaction between melody/accompaniment section and direction of influence. In the plots, it is clear that the primo-to-secondo influence is not consistently higher than the secondo-to-primo influence in passages where primos were playing the melody. Likewise the secondo-to-primo influence is not consistently higher than the primo-to-secondo influence in passages where secondos were playing the melody. Rather, instances of heightened primo-to-secondo and secondo-to-primo influence seem to be interspersed throughout the performances.

\subsubsection{Coordination of gesture acceleration}
\begin{figure*}
\begin{center}
\includegraphics[width=.6\textwidth]{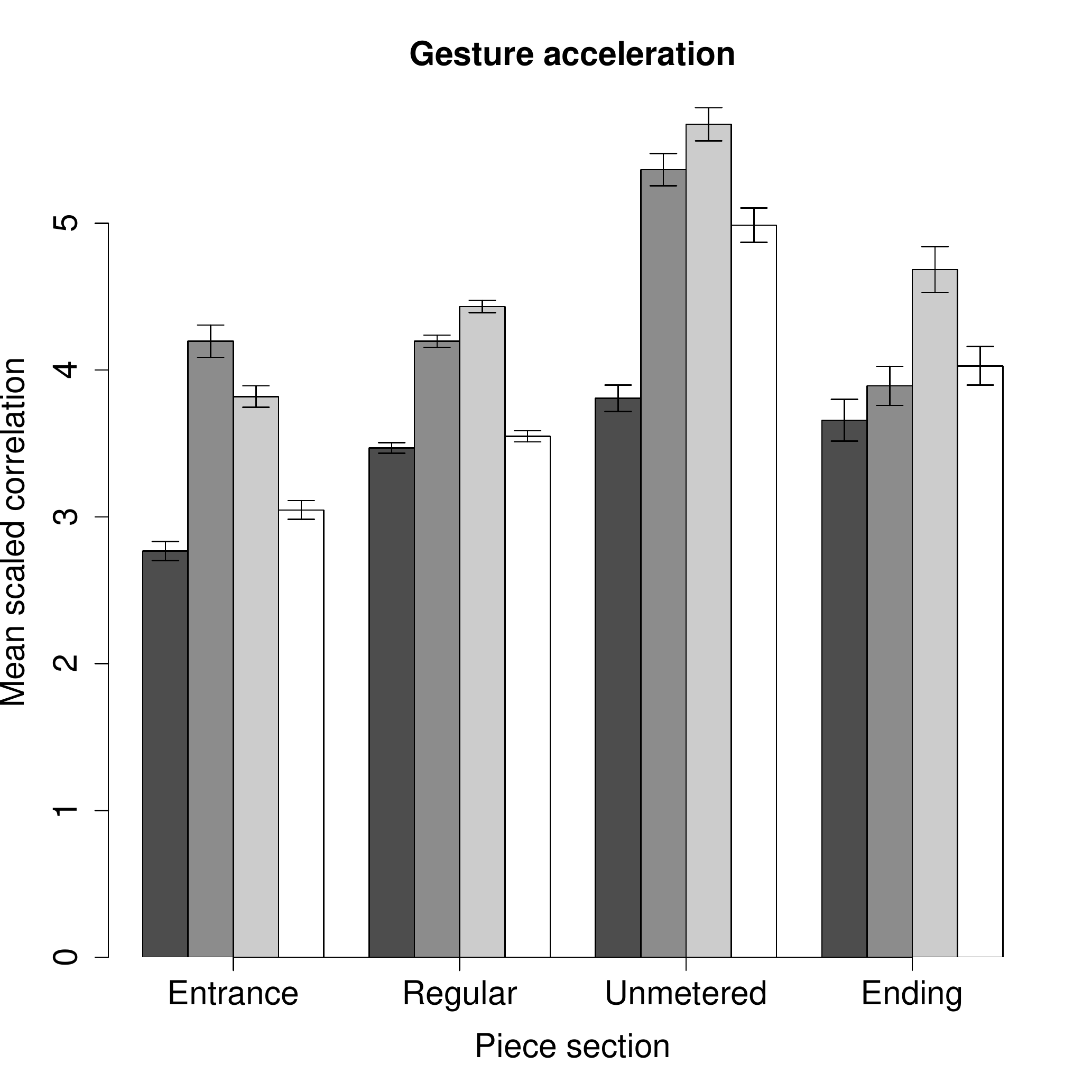}
\caption{Mean scaled correlation coefficients at lag 0, across piece sections and performances, for gesture acceleration. Error bars indicate standard error.} 
\label{fig:coordination}
\end{center}
\end{figure*}

We tested the hypothesis that coordination would increase during periods of high temporal irregularity and following rehearsal. Mean (scaled, lag 0) cross-correlation coefficients are shown by piece section and performance in Figure \ref{fig:coordination}. LME indicated significant effects of piece section, $F(3,161)=6.00,p<.001$, and performance number, $F(2,35)=9.45,p<.001$. The interaction was not significant ($p>.05$).

For piece section, post-hoc tests showed that scaled correlations were higher during the unmetered section than during the entrance, $t(161)=4.1,p<.001,\eta^2=.10$. For performance number, scaled correlations were higher during the second performance (marginally), $t(35)=2.8,p=.02,\eta^2=.05$, and third performance, $t(35)=4.3,p<.001,\eta^2=.10$, than during the first. The second and third performances did not differ.

A t-test showed that scaled correlations were higher in the third performance than in the fourth, $t(70)=3.2,p=.002,\eta^2=.13$, indicating an effect of visual contact.

Thus, taking quantity of movement into account, coordination in head movement tended to increase in the unmetered section of the piece, supporting the hypothesis that movement coordination increases during passages with less temporal stability. Coordination also improved with rehearsal, then declined again when performers were unable to see each other.

\begin{landscape}
\begin{figure}
\centering
\vspace{-1cm}
	\subfloat[First performance. \vspace{-1cm}]{{\includegraphics[width=.48\linewidth]{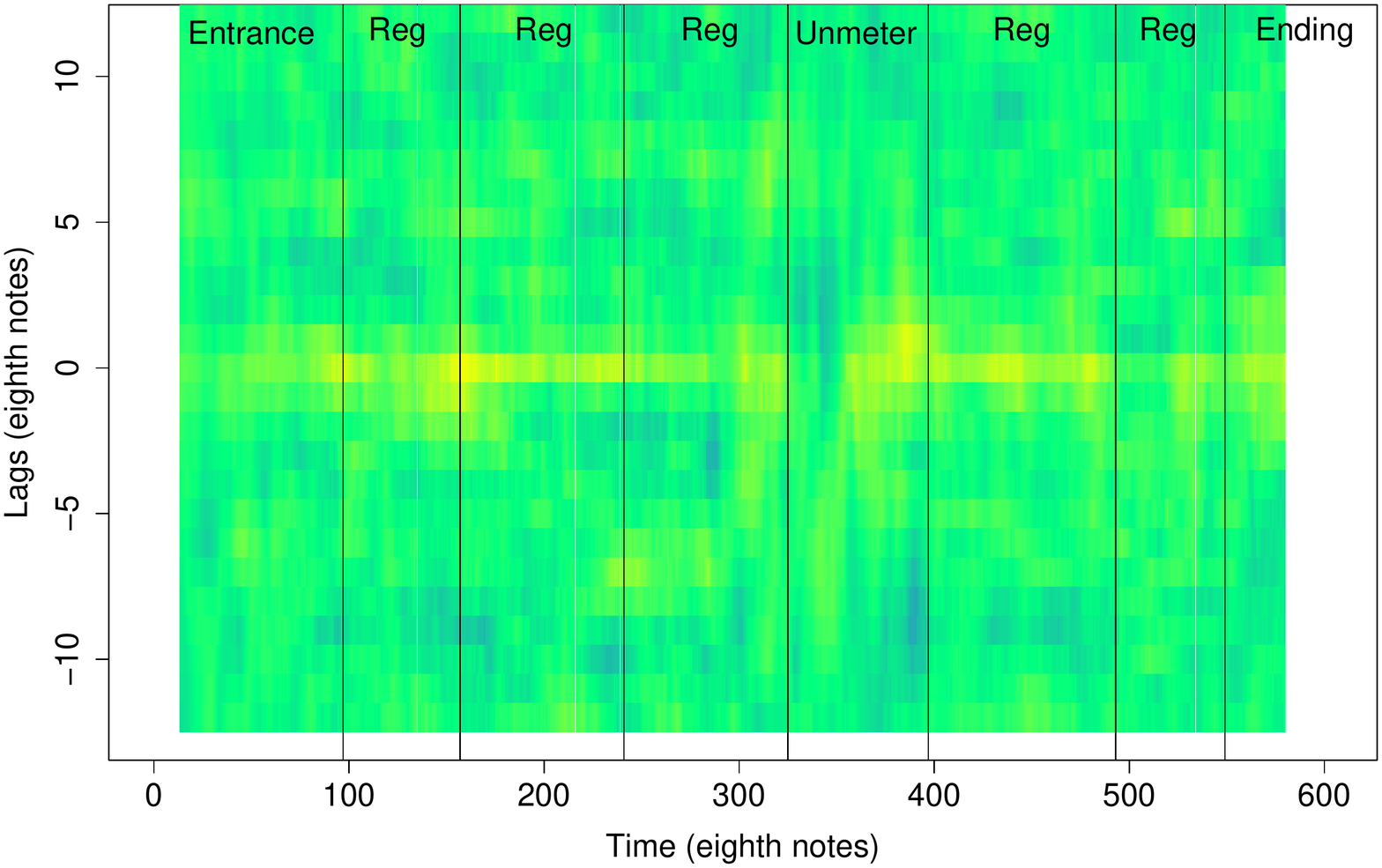}}}%
	\qquad
	\subfloat[Second performance. \vspace{-1cm}]{{\includegraphics[width=.48\linewidth]{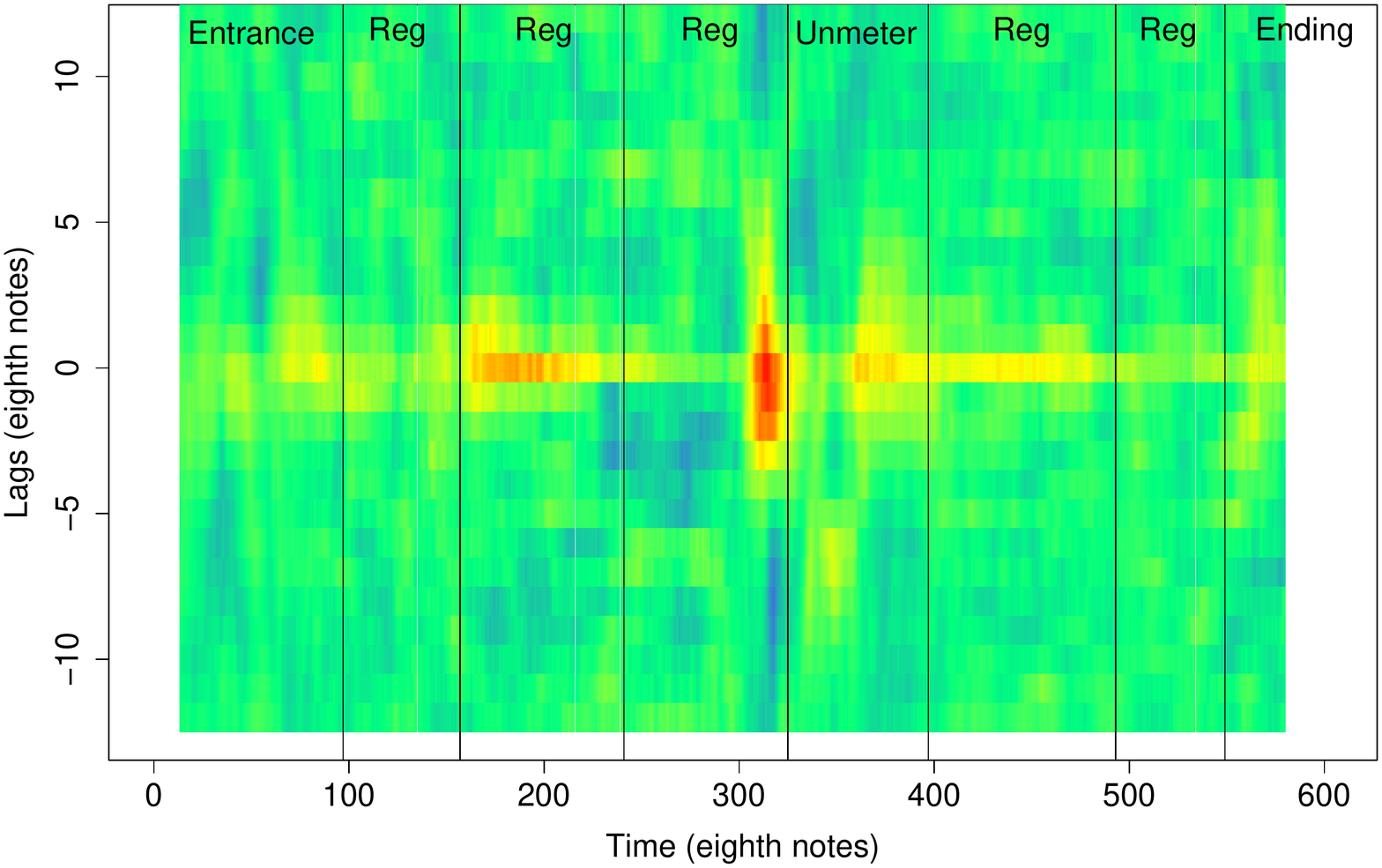}}}%
	\qquad
	
	\vspace{-1cm}
	\subfloat[Third performance. \vspace{-1cm}]{{\includegraphics[width=.48\linewidth]{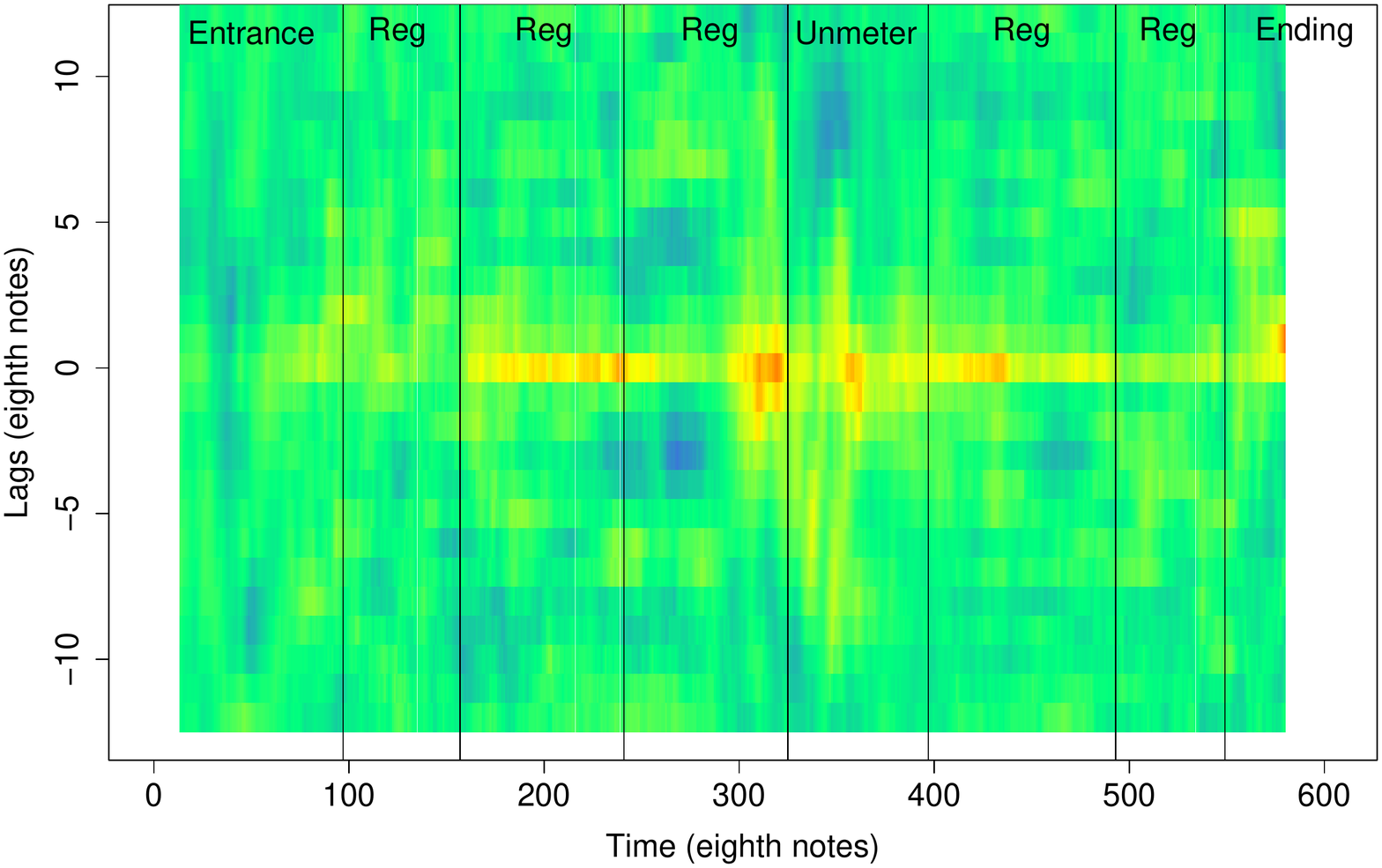}}}%
	\qquad
	\subfloat[Fourth performance. \vspace{-1cm}]{{\includegraphics[width=.48\linewidth]{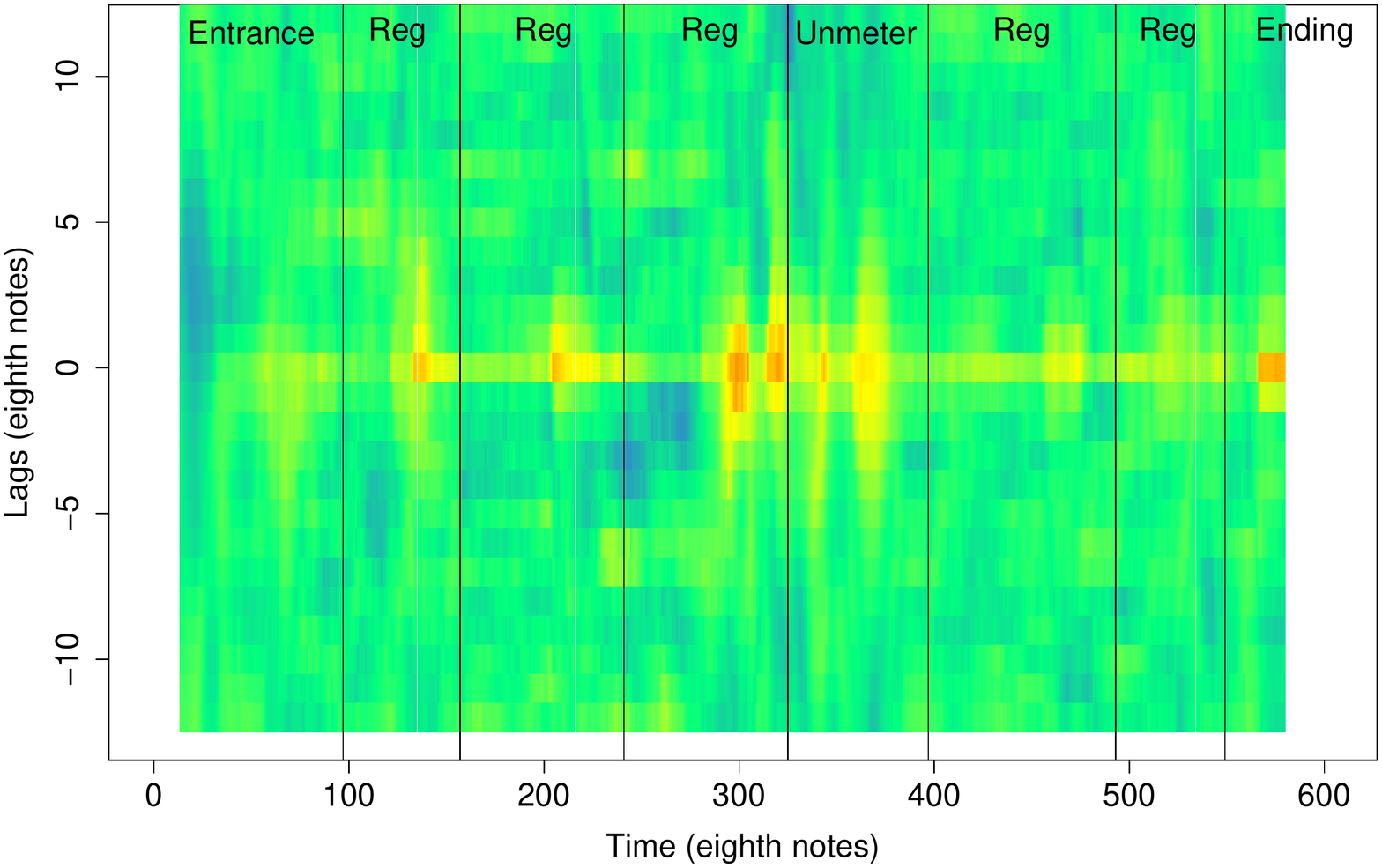}}}%
	\qquad
	\vspace{-.5cm}
\caption{Coordination in head acceleration between duo partners, averaged across duos, for performances 1--4. The heatmaps show scaled cross-correlation coefficients per eighth note across +/- 12 lag windows. X-axis values correspond to the middle note of each window. Scaled cross-correlations were computed within duos, then averaged across duos. Mean scaled correlations tended to be positive, so the plotted values range from -4.5 (blue) to +8.5 (red). Black vertical lines indicate section boundaries, and sections are labelled (``reg'' indicates regular and ``unmeter'' indicates unmetered).} \label{fig:heatmap}
\end{figure}
\end{landscape}

Heatmaps showing signed scaled cross-correlation coefficients across +/- 12 lag windows for head acceleration are presented in Figure \ref{fig:heatmap}. The values in these plots were obtained by computing the scaled cross-correlations \emph{within duos}, then averaging (signed) values \emph{across duos} at each lag/window combination. Some qualitative observations can be made from these plots. First, for all performances, average correlations were biased towards positive values, indicating a broad tendency for stronger in-phase rather than anti-phase coordination. Uptakes in coordination values can be seen close to lag 0 throughout the performances, indicating that performers' coordinated movements were closely aligned in time, rather than consistently some beats apart. While correlations were strongest, on average, in the unmetered section, a noticeable increase also occurred consistently in performances 2--4 towards the end of the third regular section, just prior to the start of the unmetered section. Possible interpretations are considered in the Discussion. 

\newpage

\section{Discussion}
This study investigated the effects of piece structure, rehearsal, and visual contact on duo musicians' gestural signalling and interaction behaviours in a score-based musical context. The goal was to identify the performance conditions that encourage communicative behaviour, including the exchange of cueing gestures, changes to movement quality (e.g., magnitude or smoothness), and fluctuations in interperformer coordination. Motion capture recordings were collected as piano and clarinet duos rehearsed a new piece. Performers' head gesture kinematics were then analysed. 

As hypothesized, similarity in acceleration profiles between performers' cueing-in gestures and their movements at held notes suggests that they exchanged cueing gestures to help with synchronizing the chords that followed these notes. Also as hypothesized, consistency of gesture patterns was greater between the second and third performances than between the first and third, suggesting that performers integrated some movement patterns into their performance routine as they rehearsed. Quantity of movement increased with rehearsal, and declined when performers' visual contact was blocked. Smoothness of movement differed according to piece structure, with movement tending to be smoother in the entrance and unmetered sections than in regularly-timed sections. Performers tended to influence each other's movements more often in the entrance section than elsewhere in the piece and, overall, secondos tended to influence primos' movements more than primos influenced secondos' movements. Interperformer coordination in head movement was strongest during the unmetered section of the piece, where temporal stability was lowest. Interperformer coordination also increased with rehearsal, and decreased when performers were unable to see each other. These findings are discussed in greater detail below.

\subsection{Cueing gestures and development of a performance routine}
It was hypothesized that the phrase structure of the unmetered passage -- in particular, the occurrence of held notes, marked with fermatas, followed by chords -- would prompt performers to communicate visually. The similarity of their head acceleration profiles at these points to the acceleration profiles of their cueing-in gestures suggests that this was the case. Gestures performed during the entrance section also proved similar to cueing-in gestures, suggesting that performers may have exchanged gestural cues in that section as well. This finding provides evidence of deliberate signalling, and is in line with prior research suggesting that the temporal ambiguity associated with long held notes prompts the exchange of cueing gestures \cite{Bishop2015}. 

It should be noted that the our analysis of cueing gestures lacked some precision, as it was carried out without regard for how performers actually distributed leader/follower roles during the unmetered section of the piece. Primos had the melody in the first half of the unmetered section and secondos had the melody in the second half, so the performer who gave the cue may have changed accordingly. However, given the difficulty associated with identifying leader/follower patterns when roles are not explicitly assigned, we considered gesture data at all held notes from all performers (primo and secondo) who gave a cueing gesture prior to piece onset, and may therefore have been looking for gestures, in some cases, from performers who assumed a follower role. Thus, our analysis may actually underestimate the occurrence of deliberate cueing behaviour among performers who assume a leader role. 

It was also hypothesized that some movement patterns would be incorporated into a performance routine as duos rehearsed. Cueing gestures, in particular, were expected to be performed predictably across the rehearsal session during periods of temporal instability. The results confirmed an increase in movement consistency across the rehearsal period, in line with previous research \cite{Williamon2002}, but did not suggest reliable differences between piece sections in terms of where movements were more or less consistent. Thus, the hypothesis that movements would be more consistent during the unmetered section (where temporal stability was lowest) than elsewhere in the piece was not supported. This finding is seems somewhat in conflict with our within-performance analysis of cueing gestures (discussed above), which suggested that cueing gestures were performed reliably during the unmetered section in at all stages of rehearsal. The analysis of cueing gestures differed from the current analysis in that here, 1) comparisons were per section rather than per beat and 2) gesture trajectory was considered instead of acceleration. Thus, it is possible that cueing gestures were performed consistently, but embedded in movement sequences that were highly variable. It is also possible that performers gave cueing gestures predictably with similar acceleration patterns, but that these followed varying trajectories across performances.

\subsection{Changes to movement kinematics signal a drive to interact}
Analyses of the quantity, smoothness, and periodicity of movement were intended to explore the potential changes that happen to gesture kinematics as playing conditions (e.g., piece structure, familiarity with the music and co-performer) change. Movements were expected to become more communicative during periods of temporal instability and across the rehearsal session, due to an increased propensity to interact.

Quantity of movement was found to increase between the first and second performances, remain stable during the third, then decline in the fourth performance, when visual contact between performers was blocked. This pattern supports that hypothesis that performers interact more via body gestures as their familiarity with a shared musical interpretation increases. It also suggests that body movements may be attenuated in situations where performers cannot see each other, perhaps due to a reduced drive to interact. 

Smoothness of movement did not change reliably across performances, but movements tended to be smoother in the entrance and unmetered sections of the piece than in the regular and ending sections, in line with our hypothesis. Previously, people have been shown to synchronize more successfully with smooth gestures than with jerky gestures \cite{Bishop2017,Woellner2012}, suggesting that musicians may manipulate the smoothness of their movements as a way of making themselves more predictable. On the other hand, movements were no more or less smooth during the fourth run than during the third, which means that whether performers had the possibility of seeing each other or not had no effect on the smoothness of their movement. If increased smoothness reflected a propensity to interact visually, it should have decreased when performers could not see each other. Thus, the increased smoothness seen in the entrance and unmetered sections might instead be attributable to factors such as differences in performers' expressive response to the music. 

\subsection{Dynamics of interperformer interactions shaped by musical structure}
Within-duo analyses of interperformer direction of influence and coordination were meant to shed light on the dynamics of movement-based interaction. We hypothesized that the leader/follower relationships that were occasionally implied by the structure of the music would affect how balanced the exchange of signals between performers was. That is, while both performers were expected to engage in a continuous cycle of reinterpreting and responding to the evolving joint musical output, one performer might sometimes dominate the interaction. The hypothesis that performers would influence their partner more when leading (playing the melody) than when following (playing the accompaniment) was not supported, however. Instead, instances of primo-to-secondo and secondo-to-primo influence were distributed across the piece. 

Across duos, there were more instances of unidirectional influence when the secondo had the melody than when the primo had the melody (though the direction of influence was not reliably secondo-to-primo). Secondo-lead sections might have differed structurally from primo-lead sections in ways that made balanced interaction less likely. Some other effects of piece structure were found: there were more instances of unidirectional influence during the entrance section than during either the regular or ending sections. The entrance section of the piece presented the main melody, first in the primo voice, then in the secondo voice, and was marked \textit{con espressione} (i.e., with expression). Many duos responded to this instruction with a \textit{tempo rubato}, or free, flexible tempo \cite<see>{Bishopsubmitted}, creating a degree of temporal ambiguity that might have encouraged one performer or the other (not necessarily the one playing the melody) to take the lead. More instances of unidirectional influence were also seen in the unmetered section, though the rates of significant g-causalities did not differ significantly from the regular and ending sections. 

A similar lack of effect of leader/follower roles was observed in \citeA{Bishopsubmitted}, where we analysed eye gaze patterns using data collected during the same duo recording sessions as reported on here. In that study, the percentage of time that performers spent watching their partner did not depend on whether they were leading or following (i.e., playing the melody versus accompaniment), in contrast to our hypothesis that they would watch their partner more when following than when leading. Our results are also in line with research showing that musicians engage in automatic and bidirectional error-correction regardless of their leader/follower assignment \cite{Himberg2017}.   

Musical structure is only one of many factors that can prompt duo performers to assume leader/follower positions, and in the case of these performances, may have been outweighed by other factors. In most previous studies where effects of leader/follower roles have been observed, leader/follower roles have been assigned by the experimenters \cite<e.g.,>{Chang2017,Kawase2014,Timmers2014} or strongly implied by the organization of the ensemble (e.g., conductor/string ensemble; see \citeA{DAusilio2012}). Here, we intended to test how leader/follower roles that arise organically and alternate throughout the performance might influence the balance of interperformer interaction. In the future, some of the social factors that affect performer relationships might be manipulated to see if they have a greater effect on movement than the structural factor considered here.

Interperformer coordination was assessed at lag 0, with the aim of quantifying the similarity between duo partners' movement patterns. Periods of strengthened coordination occurred in the unmetered section, suggesting that performers may coordinate their movements more in passages with low temporal stability. This increased coordination could serve to strengthen the perceptual-motor coupling that occurs between performers by providing more salient visual signals, better enabling them to accommodate fluctuations in each other's output.

Coordination strengthened with rehearsal, with a substantial jump in scaled correlation values occurring between the first and second performances. This increase supports our hypothesis that familiarity with a shared musical interpretation (and more broadly with the music and co-performer) encourages performers to interact via their body movements. (Note that the design of the current study does not allow us to distinguish between the effects of ``musical'' familiarity (i.e., with the notes and shared interpretation) and ``social'' familiarity (i.e., with the co-performer); this point should be addressed in the future.) The decline in coordination strength observed between the third and fourth performances indicates that movement coordination was dependent on performers' ability to see each other, and provides further evidence that the strengthened correlations observed during the second and third performances reflect visual interaction, rather than similar responses to the music.

The generalizability of these findings are, of course, limited by the narrow focus of this study. We examined the interaction dynamics that arise in the context of \emph{score-based} music performance. Most musical traditions are not score-based. Moreover, different musical traditions have different constraints and expressive conventions, so uncertainty may arise from different sources than is the case for Western art music. In particular, the challenge of maintaining synchronization during periods of temporal irregularity -- while a primary concern for Western classical musicians -- may be less relevant when the music is beat-based. In such cases, uncertainty might relate primarily to which harmonic progressions or rhythmic patterns are used. The conditions that encourage visual interaction might differ from the Western art music context as a result.

\subsection{Conclusions}
Research has previously shown that small ensembles do not have to see each other in order to coordinate temporally. Indeed, \citeA{Bishopsubmitted} found no effects of visual contact on note synchronization in the recordings reported on here. The current study, however, has shown that performers' ability to see each other can have other effects -- influencing the strength and quality of interperformer interactions, for instance.

This study sought evidence of movement-based interaction between performers; however, it could be argued that some of our findings are not necessarily evidence of interperformer interaction, but rather evidence of commonalities in patterns of expressive movement. For example, performers may move little when in the earliest stages of learning a new piece because they have not yet decided on how the piece should be interpreted and are focused on playing the correct notes. The increased quantity and consistency of movement patterns with rehearsal, thus, could simply reflect the development of an expressive, embodied interpretation. On the other hand, we would argue that the effects of visual contact on quantity of movement and interperformer coordination provide evidence that increased movement is at least partially driven by a desire to interact. Furthermore, the increased rate of significant unidirectional Granger causality tests during the entrance section of the piece (and to a lesser extent, the unmetered section) provides evidence of performers influencing each other's movement patterns.

The results presented here show evidence that signalling as well as interaction processes are involved in a successful duo performance. The increase in consistency of movements that occurred with rehearsal suggests that performers gradually establish a performance routine. Performers were found to exchange cueing gestures at held notes across performances, indicating that practiced communicative movements can be a part of this routine. 

Our findings raise an important question: what function might communicative and interactive movement features serve, given that they are largely unnecessary for temporal coordination? Given the differences in gesture kinematics that we observed between piece sections, we might hypothesize that, in passages with low temporal stability, movement-based interaction supports automatic coordination/synchronization mechanisms by affirming planned tempo fluctuations, perhaps increasing performers' confidence and freeing up cognitive resources for other processes (e.g., considering new interpretive options). 

More broadly, our observations of duos' movement and gaze behaviour suggest that performers choose to interact visually when they have the opportunity and means to do so. We would therefore hypothesize that visual interaction serves as a social motivator, which may affect performance success by encouraging creative thinking and risk-taking, while also enhancing performers' enjoyment and absorption in the performance task. The visual modality may provide performers with a means of exchanging information that is not communicated adequately through their musical sound signals (e.g., information relating to attention and engagement). This supplementary layer of communication might help performers focus their attention on the shared task, establish a mutual awareness of each other's contributions, and minimize feelings of self-consciousness (facilitated by co-performers' constant feedback). These factors are thought to underlie the peak levels of ``togetherness' that characterize periods of group flow \shortcite<see>{Cochrane2017,Gaggioli2017,Hart2014}.

Future research might investigate the role that social motivation plays in performance success, and how interperformer interaction helps to maintain it. As mentioned earlier, performers' expressive body movements are informative to both co-performers and audiences, and many of the variations in movement patterns that could be categorized as interactive at an interperformers level may also shape audiences' perceptions of the performance. Further study is needed to show how movement-based interperformer interaction affects audience members' experiences, including their perceptions of ensemble coordination, coherence, and expressivity.

\subsection*{Acknowledgements} 
This research was supported by Austrian Science Fund (FWF) grant P29427, and European Research Council (ERC) grant 670035 (project ``Con Espressione''),  under the EU's Horizon 2020 Framework Programme. We are grateful to Manuel Gangl for his help with arranging the duet for clarinets, and to Ayrin Moradi, Anna Pudziow, and Martin Bonev for their assistance with participant recruitment and data collection.

\bibliographystyle{apacite}
\bibliography{MadPig_ref}

\newpage
\appendix
\Large \textbf{Appendix} \normalsize
\section{Cueing gestures}
\subsection{Identification of cueing-in gestures at piece onset}
Prior research has shown that cueing-in gestures are characterized by a local maximum in acceleration, followed by a local minimum. These instances of rapid deceleration communicate beat position, while the magnitude of acceleration change relates positively to the clarity of communicated beats \cite{Bishop2017,Bishop2018,Luck2008a,Luck2009}.

In line with these findings, cueing-in gestures in the current study were defined as occurring where a local maximum in acceleration (8 consecutively increasing samples followed by 8 consecutively decreasing samples) was followed by a local minimum (3 consecutively decreasing samples followed by 3 consecutively increasing samples). The local maximum had to occur within the one-beat interval prior to piece onset (where 1 beat = duration of the first performed quarter note beat), and had to be above the midway point between the highest and lowest acceleration values recorded during the two-beat interval surrounding piece onset.

It should be noted that very subtle gestures might not pass this final criterion if the performer were to start moving more substantially immediately following piece onset (i.e., during the first performed beat interval), increasing the average acceleration value for the two-beat period. Such gestures do not meet our definition of cueing-in gesture, as they cannot be distinguished on the basis of acceleration. For these performances, other audio/visual cues, such as facial expressions or breathing, may play a greater role in communicating piece onset than does head acceleration.

\begin{figure}[h]
\centering
	\subfloat{{\includegraphics[width=\textwidth]{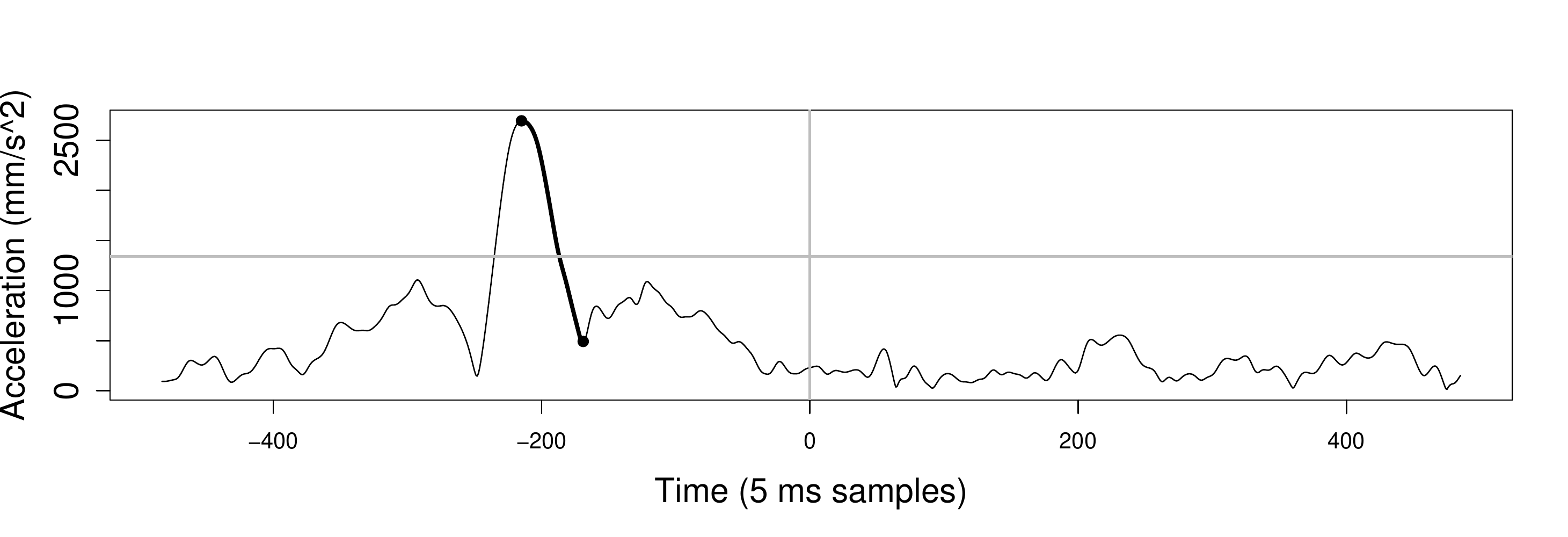}} 
	\label{fig:cueing_in1}}%
	\qquad
	
	\vspace{-1cm}
	\subfloat{{\includegraphics[width=\textwidth]{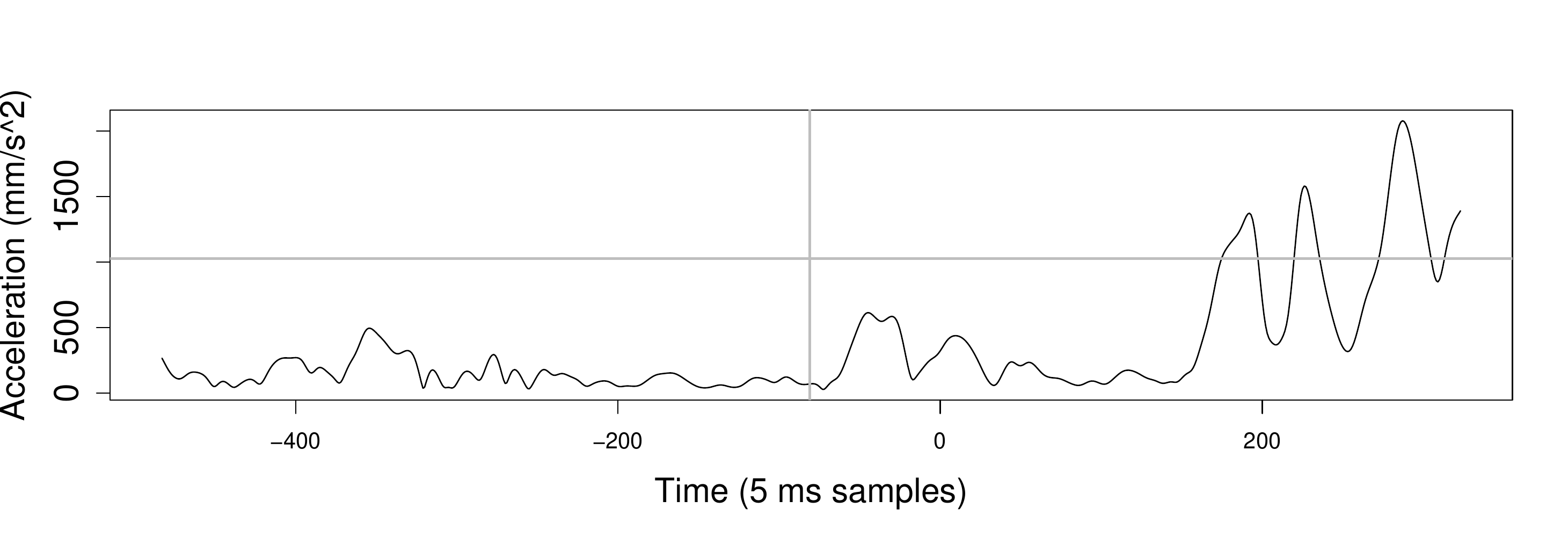}} 
	\label{fig:cueing_in2}}%
	\qquad
\caption{Sample acceleration curves showing the presence (top) and absence (bottom) of a cueing-in gesture. Vertical grey lines indicate piece onset; horizontal grey lines indicate the midway point between the maximum and minimum acceleration values in the 2-beat interval. Acceleration maxima had to be above this point for a cueing-in gesture to be identified. The bold segment of the acceleration curve corresponds to the characteristic cueing-in gesture deceleration.}
\label{fig:cueing_in}
\end{figure}

An example of an acceleration curve containing a cueing-in gesture is shown in Figure \ref{fig:cueing_in}.

\subsection{Identification of cueing gestures throughout performances}
For performances in which a cueing-in gesture could be identified, a within-performance search was made for similar gestures: using dynamic time warping, we estimated the similarity between the acceleration curves given during the one-beat (quarter note) interval before piece onset (i.e., the ``reference'' segment containing the cueing-in gesture) and acceleration curves given during each subsequent quarter note in the performance (``test'' segments). Beats were counted at the quarter note level since the entrance section of the piece had a 6/4 meter. (While some performers may have conceptualized the meter differently (e.g., as 3/2), the majority of performances did contain a cueing in gesture during the quarter note beat period before piece onset.)

Alignments were made using a symmetric, normalizable step pattern, and constrained to a slanted band window with a width of 100 observations. One-beat test segments were aligned to the reference segment in their entirety; partial alignments were not permitted. A normalized distance value (i.e., indicating average distance per step, accounting for differences in section lengths) was obtained for each test segment that indicated its similarity to the reference segment containing the cueing-in gesture.

\section{Consistency}
\begin{figure}
	\subfloat{{\includegraphics[width=.5\textwidth]{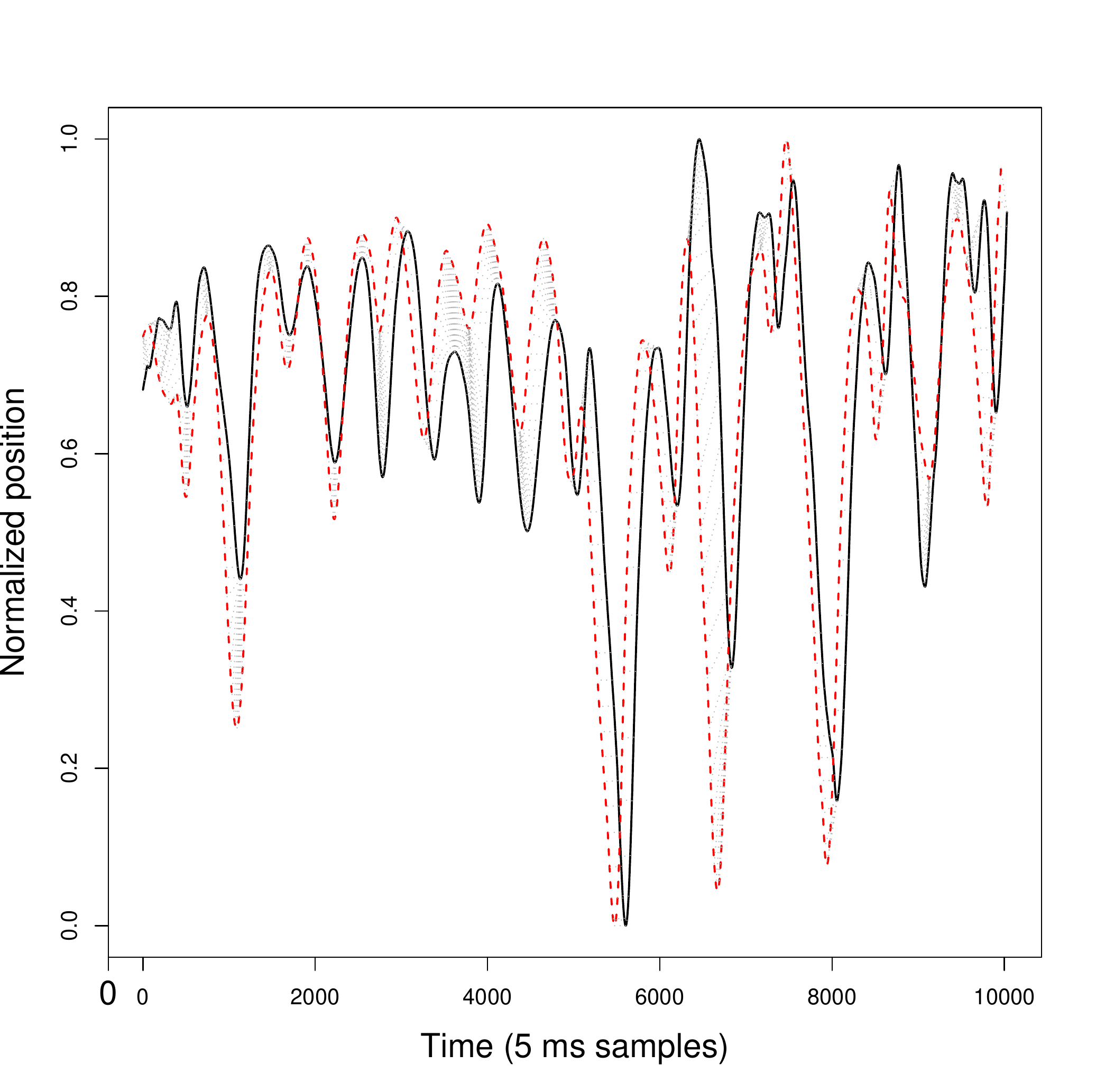}}}%
	\qquad
	\subfloat{{\includegraphics[width=.5\textwidth]{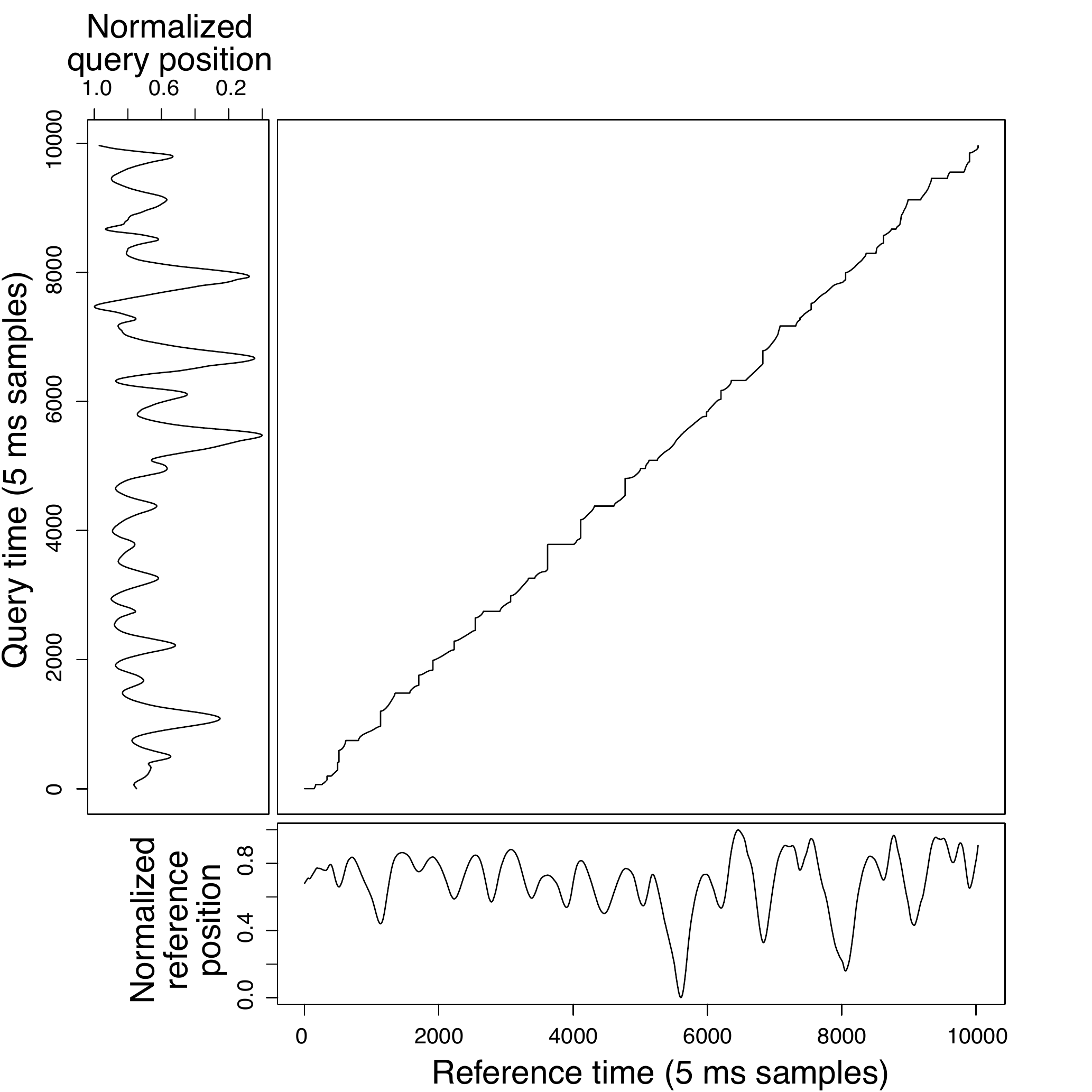}}}%
	\qquad
	
	\vspace{-.3cm}
	\subfloat{{\includegraphics[width=.5\textwidth]{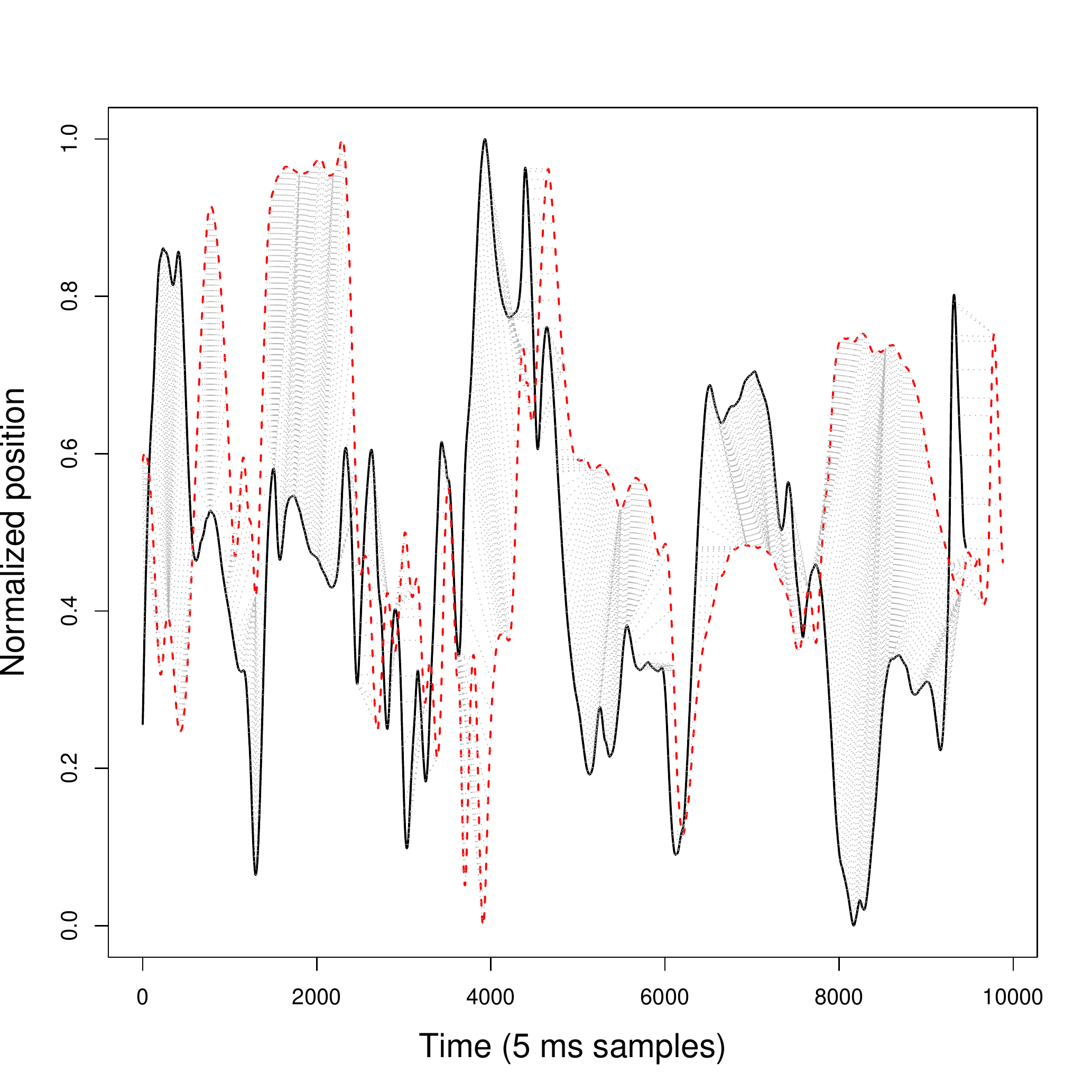}}}%
	\qquad
	\subfloat{{\includegraphics[width=.5\textwidth]{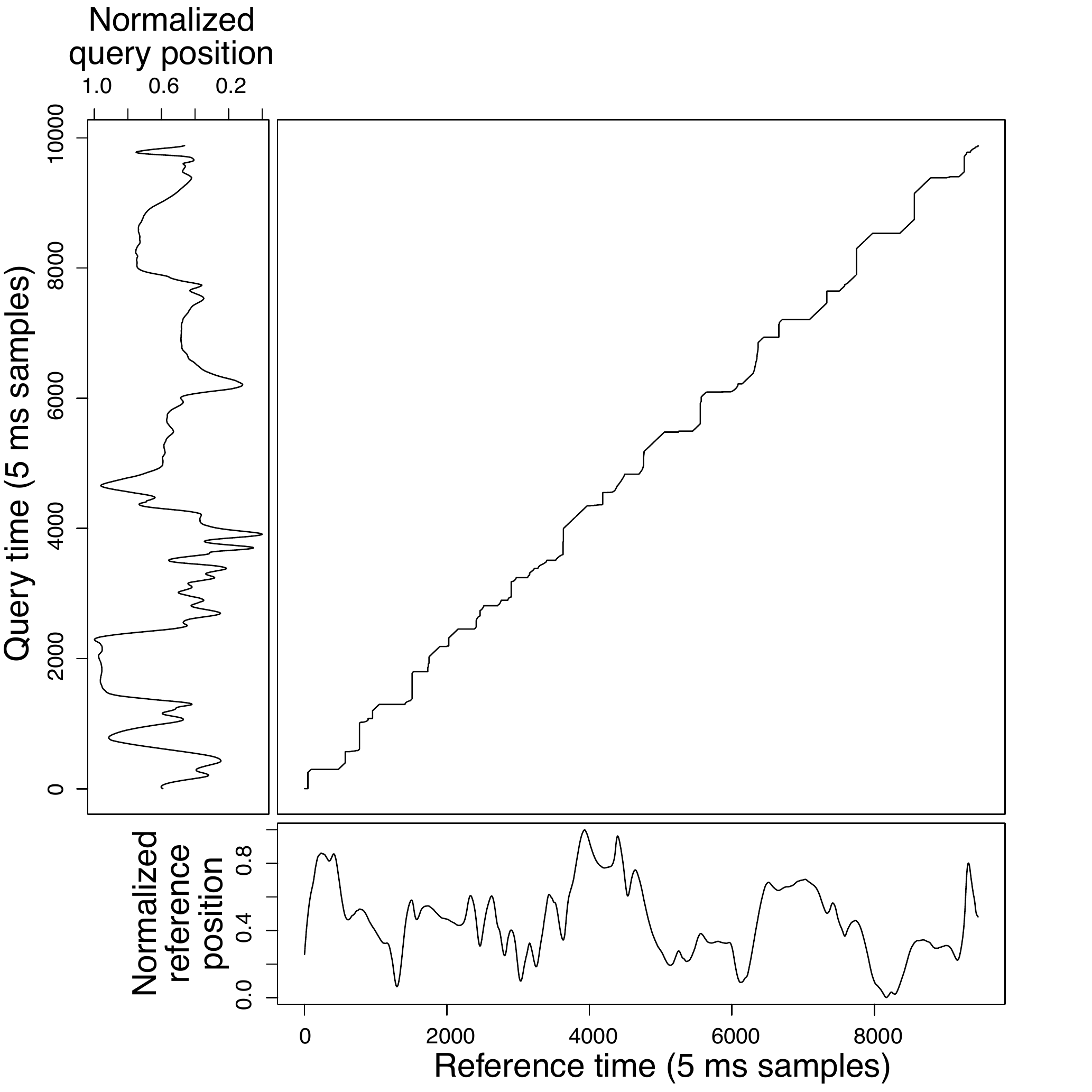}}}%
	\qquad
	\caption{Examples of dynamic time warping of (top) a pianist's 2nd and 3rd performance entrance sections, and (bottom) a clarinettist's 1st and 3rd performance unmetered sections. The two-way plots on the left show how individual points on test series (black solid lines) were mapped to corresponding points on reference series (red dotted lines). The three-way plots on the right show the warping paths that were mapped out between the test and reference series. The normalized distances between test and reference series were (top) .023 and (bottom) .175.} \label{fig:dtw}
\end{figure}

Comparisons between performance sections were made using dynamic time warping. Gesture trajectories in test and reference performance sections were first normalized to a range of 0 -- 1. Each test section was then aligned to the reference using a symmetric, normalizable step pattern. The alignment was constrained to a slanted band window with a width of 200 observations. Sections were aligned in their entirety; partial alignments were not permitted. Normalized distances between the aligned test and reference sections were taken as an indication of within-subject/between-performance consistency (Figure \ref{fig:dtw}).

\section{Quantity of motion}
Quantity of motion, as stated in section 2.5.2, was calculated as the mean absolute distance travelled (in m) per 5 ms observation for each performance section; that is, as the sum of absolute differences between successive observations, divided by the total number of observations. For a variable comprising 3D position values $pos_1, pos_2, ..., pos_n$,

\begin{equation*}
\text{Quantity of motion}= \frac{1}{n}\sum\limits_{i=2}^{n}\left|pos_{i}-pos_{i -1}\right|
\end{equation*}

\vspace{-2em}
\section{Direction of influence}
Direction of influence was evaluated for head acceleration using Granger Causalities. First, motion data were resampled by averaging per eighth note. An eighth note beat interval was chosen because this metrical level was common to all meters in the piece (i.e., all meters could be counted or subdivided into eighth notes -- this would allow us to compute g-causalities over a rolling window across the entire piece.

Granger Causality estimates the likelihood that data series $x$ influences data series $y$ by comparing two models: a restricted model in which observations of $y$ are predicted only by lags of $y$, and an unrestricted model in which observations of $y$ are predicted by lags of $x$ in addition to lags of $y$. The predictive ability of the two models is then compared. We computed g-causalities (using the ``lmtest'' package in R) for a rolling window of 30 eighth note lags across each performance. Models were specified to include one predictive lag, since between-performer coordination generally seemed strongest at one lag (discussed in section 3.3.2). The g-causality test that we used compares restricted and unrestricted models using a Wald test, and a significant result suggests a difference in their predictive ability.

\section{Between-performer coordination}
Just as for the Granger Causality analyses described above, motion data were averaged per eighth note. Cross-correlations were then computed across a rolling window of 25 eighth note lags across each pair (primo/secondo) of performances.

These cross-correlation values were scaled to account for variability in quantity of motion. This was necessary because windows in which both primo and secondo moved very little could lead to seemingly high correlations. Separate quantity of motion calculations were made for each performer in each window, using the equation given for ``Quantity of motion'' above, then averaged between partners to give a single measure per duo, per window. This mean ``per-window'' quantity of motion was then taken as a fraction of the global mean quantity of motion (computed for the entire performance and averaged between partners). The resulting fraction was used to scale lag 0 cross-correlation values, which were subsequently averaged per piece section.
\end{document}